# Strain-Gradient and Curvature-Induced Changes in Domain Morphology of BaTiO$_3$ Nanorods: Experimental and Theoretical Studies

Olha A. Kovalenko[1,2,3] *, Eugene A. Eliseev[2], Yuriy O. Zagorodniy[2], Srečo Davor Škapin[1], Marjeta Maček Kržmanc[1], Lesya Demchenko[4,5], Valentyn V. Laguta[2,6], Zdravko Kutnjak[1], Dean R. Evans[7]†, and Anna N. Morozovska[8]‡

[1] *Slovenia Jožef Stefan Institute, Ljubljana, Slovenia*

[2] *Frantsevich Institute for Problems in Materials Science, National Academy of Sciences of Ukraine, 3, str. Omeliana Pritsaka, 03142 Kyiv, Ukraine*

[3] *LLC NanoTechCenter, Kyiv, Ukraine*

[4] *Stockholm University, Department of Chemistry, Sweden*

[5] *Ye. O. Paton Institute of Materials Science and Welding, National Technical University of Ukraine "Igor Sikorsky Kyiv Polytechnic Institute"*

[6] *Institute of Physics of the Czech Academy of Sciences, Na Slovance 1999/2, 18200 Prague 8, Czech Republic*

[7] *Air Force Research Laboratory, Materials and Manufacturing Directorate, Wright-Patterson Air Force Base, Ohio, 45433, USA*

[8] *Institute of Physics, National Academy of Sciences of Ukraine, 46, pr. Nauky, 03028 Kyiv, Ukraine*

**Abstract**

We investigate the impact of OH$^-$ ions incorporation on the lattice strain and spontaneous polarization of BaTiO$_3$ nanorods synthesized under different conditions. It was confirmed that the lattice strain depends directly on Ba supersaturation, with higher supersaturation leading to an increase in the lattice strain. However, it was shown that crystal growth and observed lattice distortion are not primarily influenced by external strain; rather, OH$^-$ ions incorporation plays a key role in generating internal chemical strains and driving these processes. By using the less reactive TiO$_2$ precursor instead of TiOCl$_2$ and controlling Ba supersaturation, the slower nucleation rate enables more effective regulation of OH$^-$ ions incorporation and crystal growth. This in turn effects both particle size and lattice distortion, leading to *c/a* ratio of 1.013 – 1.014. The incorporation of $OH^-$ ions induces lattice elongation along the c-axis, contributing to anisotropic growth, increasing of the rod diameter and their growth-induced bending. However, the possibility of the curvature-induced changes in domain morphology of $BaTiO_3$ nanorods remains almost unexplored. To study the possibility, we perform analytical calculations and finite element modeling, which provide insights into the curvature-induced changes in the strain-gradient, polarization distribution, and domain morphology in $BaTiO_3$ nanorods. Theoretical results reveal the appearance of the domain stripes in $BaTiO_3$ nanorod when the curvature exceeds a critical angle. The physical origin of the domain stripes emergence is the tendency to minimize the elastic energy of the nanorod by the domain splitting. These findings suggest that $BaTiO_3$ nanorods, with curvature-controllable amount of domain stripes, could serve as flexible race-track memory elements for flexo-tronics and domain-wall electronics. Overall, this work enhances the understanding of how the shape anisotropy, lattice strains, and strain gradients influence the domain morphology of ferroelectric nanorods, offering a pathway for tuning properties of the nanorods for advanced applications in nanoelectronics.

---

* Corresponding author, e-mail: olgiuskovalenko@gmail.com

† Corresponding author, e-mail: dean.evans@afrl.af.mil

‡ Corresponding author, e-mail: anna.n.morozovska@gmail.com

## I. INTRODUCTION

It is well known that lattice strain induces remarkable enhancement in the piezoelectric and ferroelectric performance of perovskite-based materials [1, 2, 3, 4, 5]. The importance of tailoring lattice strain in halide perovskite crystals makes strain engineering a powerful tool for enhancing the performance of energy conversion or storage devices. For instance, applying a compressive strain of -2.3% or more (in magnitude) to $BaTiO_3$ reduces its bandgap, transforming it into a direct bandgap semiconductor; eliminating additional conduction band interactions is a notable advancement for photovoltaic devices [3].

Curvature control of the polarization distribution and domain structure morphology in ferroelectric single-crystals [6], layered 2D ferrielectrics [7, 8], buckled ferroelectric free sheets [9], and adaptive ferroelectric membranes [10] recently became of significant interest due to the insight to the strain-gradient and curvature-induced changes of the ferroelectric long-range order, dipole-dipole correlations, domain wall structure, and electronic properties [11]. The curvature-controllable domain stripes are highly promising as flexible race-track memory elements for flexo-tronics and domain-wall electronics.

However, the possibility of strain-gradient and/or curvature-induced changes in the domain morphology of $BaTiO_3$ nanoparticles and nanosheets remain largely unexplored, although the ferroelectric material is highly sensitive to various kinds of strains. For instance, Choi et al. [12]

demonstrated experimentally that with epitaxial stress (lattice mismatch) the spontaneous polarization $P_s$ of BaTiO$_3$ films increased 250 %, while the Curie temperature $T_C$ increased to approximately 500°C. First principles calculations of Ederer and Spaldin [13] and phenomenological thermodynamic theory [14] corroborate the experimental results. Subsequently, it was shown that the chemical strains are responsible for the strong increase in the Curie temperature (above 167°C) and tetragonality (up to 1.032) near the surface of a BaTiO$_3$ film with injected oxygen vacancies [15]. The lattice constants mismatch, which induces epitaxial strains and/or strains between the core and shell, are responsible for the preservation and enhancement of $P_s$, tetragonality $c/a$ and $T_C$ increase in BaTiO$_3$ core-shell nanoparticles, as it was found in Refs. [16, 17, 18, 19]. The structural origin of recovered ferroelectricity and the phenomenological description of this striking effect in given by Zhang et al. [19] and Eliseev et al. [20], respectively.

In this work, we explore the impact of strain-gradients and curvature on the domain morphology of BaTiO$_3$ nanoparticles using analytical calculations and finite element modeling (FEM). These methods provide insights into the strain gradient-induced and curvature-induced changes of the polarization distribution and domain morphology in the BaTiO$_3$ nanorods.

## II. EXPERIMENT
### A. Materials and Methods

BaTiO$_3$ nanorods were synthesized using a single-step hydrothermal technique similar to that reported by Inada [21] and Kovalenko [22]. However, different Ti$^{4+}$ sources (TiO$_2$ and TiOCl$_2$ solution) and different concentrations of BaCl$_2$ and NaOH were used. Nanopowder TiO$_2$ (P25, 99/00–100.5 Evonik), along with TiCl$_4$ (≥99.0% (AT), Fluka), BaCl$_2$·2H$_2$O (<99%, Carlo Erba), ethylene glycol (≥99%, Sigma-Aldrich), and NaOH (98.7%, Fisher Scientific) were used as the initial reagents. To obtain a 1 M TiOCl$_2$ solution, 15 mL of the TiCl$_4$ solution (concentration ≥99.0%) was slowly added to a flask containing ice water under a fume hood to mitigate the exothermic reaction and prevent the release of toxic HCl vapor.

To obtain the samples S1 and S3, 0.001 mol of TiO$_2$ nanopowder was added to 20 ml of 0.2-M and 0.3-M solution of BaCl$_2$, respectively, under continuous stirring at 200 rpm at room temperature. After this, 10 ml of 10-M NaOH (Na/Ba ratio was 33) was added dropwise to the solution, followed by the addition of 5 ml of EG and 15 ml of water. Finally, the prepared mixture was placed into a Teflon-lined stainless-steel autoclave (100 ml) with a half-filled volume and transferred to the furnace for hydrothermal treatment. The reaction was carried out at a temperature of 200 °C and a pressure of 10 bar for 24 hours. The precipitation was decanted using an Eppendorf 5810R centrifuge and washed with distilled water and acetic acid to remove the residual barium carbonate (BaCO$_3$). The wet powder was dried in an oven at 50 ºC for 12 hours. Sample S2 appeared to be quasi-spherical $BaTiO_3$ nanoparticles with an asymmetric structure attributed to a mixture of cubic and tetragonal phases. This sample will not be considered elsewhere. Samples S4 and S5 were obtained in the same way as samples S1 and S3, except that $TiOCl_2$ solution was used as the $Ti^{4+}$ source instead of $TiO_2$. The $Ba^{2+}$ concentration for samples S4 and S5 was 0.3 M and 0.2 M, respectively. Moreover, sample S5 was obtained at a Na/Ba ratio of 10 instead of 33.

The phase and structure of the as-prepared $BaTiO_3$ nanopowder were determined by X-ray diffraction (XRD) using a Bruker AXS D4 ENDEAVOR diffractometer with Cu–Kα radiation. The crystallite size was evaluated based on the size of the coherent scattering region calculated using the Scherrer equation and the full width at half-maximum (FWHM) of the (100) and (001) diffraction peaks. The tetragonality ($c/a$) was determined from the (200) and (002) reflection. The broadening of the low-angle diffraction peaks was also used to estimate the particle size, strains, and anisotropy using the Williamson-Hall technique. The sizes of the $BaTiO_3$ nanorods were analyzed using a field-emission scanning electron microscope (FESEM) (Zeiss ULTRA plus) employing a voltage of 3 kV. The size distribution was obtained from SEM images by manually counting the populations of 500 nanoparticles using ImageJ and Origin software.

The nuclear magnetic resonance (NMR) spectra of samples S1-S5 were recorded in a Bruker Avance II 400-MHz commercial NMR spectrometer at a magnetic field of 9.40 T, corresponding to the Larmor frequency of 44.466 MHz. The $^{137}Ba$ NMR spectra were obtained with the conventional 90x−τ−90y−τ spin echo pulse sequence using a four-phase "exor-cycle" phase sequence (xx, xy, x–x, x–y) to form echoes with minimal distortions due to anti-echoes, ill-refocused signals, and piezo-resonances. The π/2-pulse length was typically $t_{\pi/2} = 1\mu s$, the spin-echo delay time $\tau$ was 20 $\mu s$ and the repetition delay between scans was 0.15 s. Due to the limited sample quantity, more than 2,000,000 scans were collected to obtain the spectrum. The spectra were measured at room temperature. The $^{137}Ba$ spectra were referenced against $BaF_2$, using its narrow $^{137}Ba$ resonance as the zero for the chemical shift.

### B. Physicochemical Properties of BaTiO₃ Nanorods

To investigate how the synthesis conditions, size, and structure of anisotropic $BaTiO_3$ nanoparticles affect their polarization distribution, strain, and possible domain morphology, we selected $BaTiO_3$ nanorods with varying diameters (50 – 90 nm), lengths (410 – 900 nm), and aspect ratios (6 – 18) (see **Table 1** and **Fig. 1**).

**Table I. Characteristics of the BaTiO₃ nanorods**

| Sample | Width* (nm) | Length* (nm) | Aspect ratio* (nm) | $L_{001}^a/L_{00L}^b$ (nm) | $L_{100}^a/L_{H00}^b$ (nm) | $\Delta a/a$ | $\Delta c/c$ | $c/a^M$ $c/a^G$ | $e_{001}$ | $e_{100}$ |
|---|---|---|---|---|---|---|---|---|---|---|
| S1 | 70 | 410 | 6 | 38/41 | 81/70 | 0.0025 | 0.0021 | $1.013^M$ $1.012^G$ | 1.84 | 0.73 |
| S3 | 90 | 770 | 9 | 42/42 | 85/73 | 0.0035 | 0.0034 | $1.013^M$ $1.012^G$ | 1.67 | 0.66 |
| S4 | 50 | 800 | 16 | 37/38 | 46/42 | 0.0031 | 0.0012 | $1.011^M$ $1.011^G$ | 1.81 | 0.76 |
| S5 | 50 | 900 | 18 | 39/40 | 67/59 | 0.0035 | 0.0002 | $1.009^M$ $1.009^G$ | 1.51 | 0.71 |

* Calculations are based on the SEM data,

$L_{001}^a/L_{100}^a$ is the crystallite size based on the coherent scattering region calculated with the Scherrer equation,

$L_{00L}^b/L_{L00}^b$ and $e_{001}/e_{100}$ are the crystallite size and lattice strain in the [001] and [100] crystal directions, respectively, calculated with the Williamson-Hall technique based on the XRD data,

$c/a^M$ and $c/a^G$ are c/a ratio calculated manually using X-peak software and the Gauss function with Origin software, respectively.

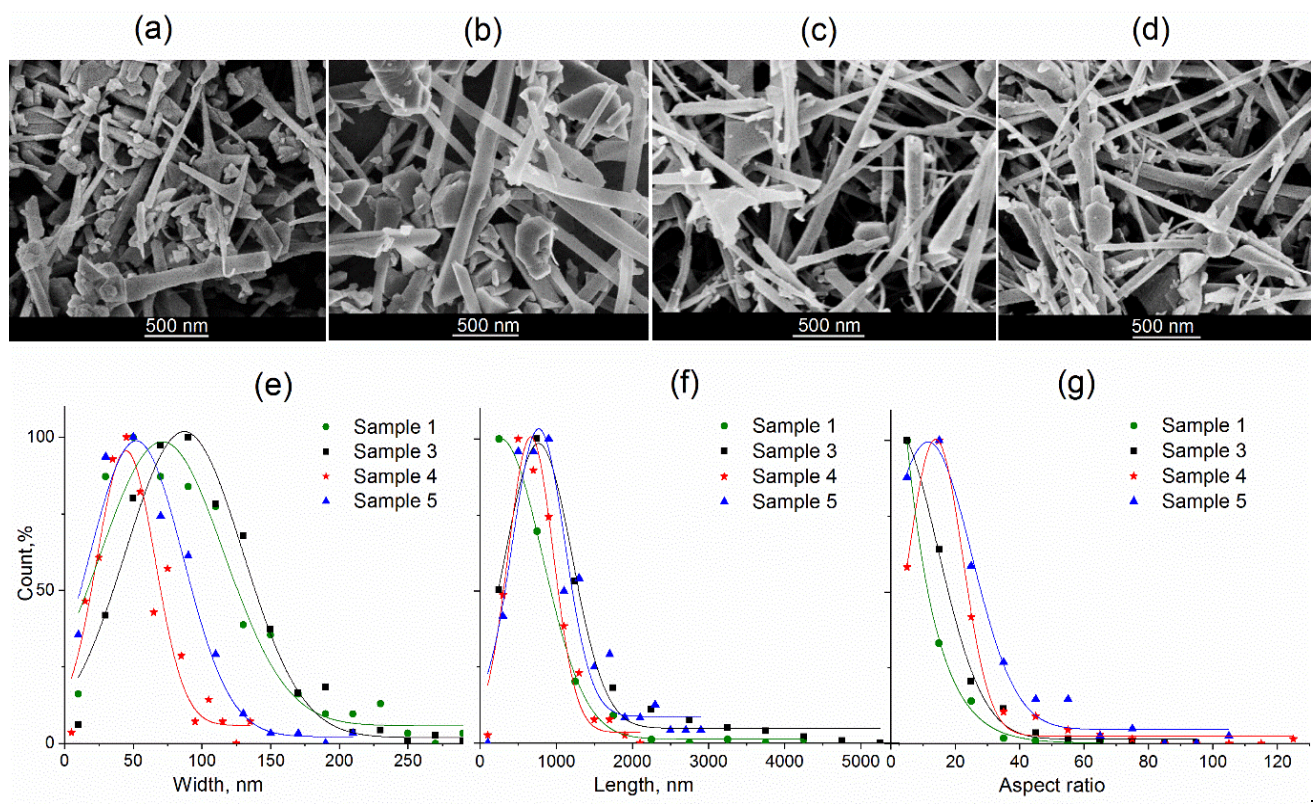

**FIGURE 1.** SEM images ((a)-(d)) of samples S1, S3, S4, and S5 showing rod-shaped BaTiO₃ nanoparticles. Particle size distribution of samples S1, S3, S4, and S5 based on SEM data: **(e)** width, **(f)** length, and **(g)** aspect ratio.

All diffraction peaks measured in these samples have been indexed to the tetragonal structure of BaTiO$_3$ (space group P4mm), which agrees with the Crystallography Open Database (COD No. 1513252); see **Fig. 2**. No traces of impurities and/or secondary phases were observed. The splitting of the XRD reflections (002) and (200) in these samples indicates the formation of a tetragonal structure. The lattice constant ratio, $c/a$, is 1.013 for samples S1 and S3, 1.011 for samples S4, and 1.009 for samples S5. The shift of the diffraction peaks (002) and (200) to lower angles in all samples, compared to bulk BaTiO$_3$, indicates lattice expansion due to the introduction of OH$^-$ ions into the crystals, apparently into trans-position [23]. This expansion leads to high tetragonality ($c/a$ ratios ranging up to 1.013), as illustrated in **Fig. 2**.

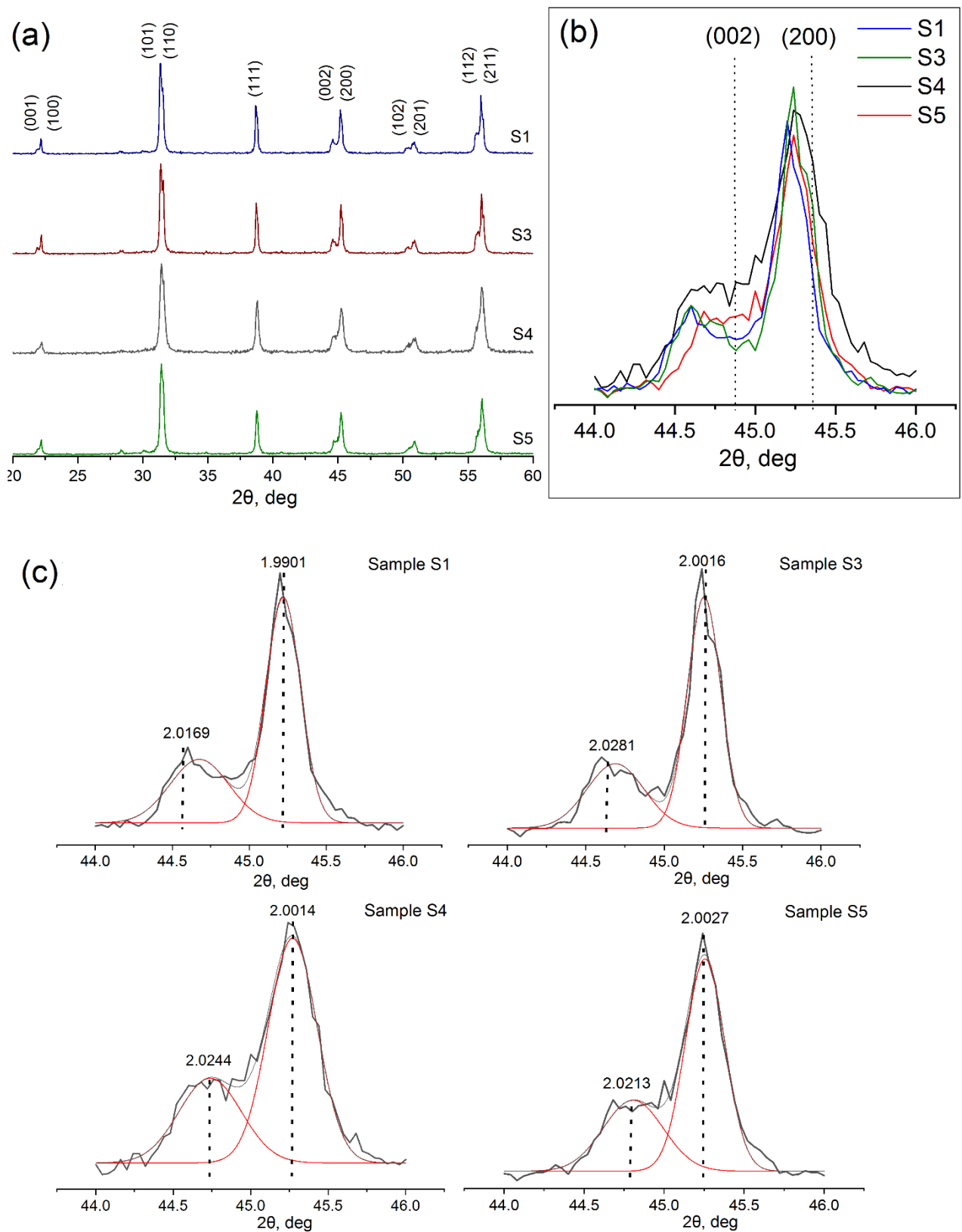

**FIGURE 2. (a)** XRD patterns of BaTiO$_3$ nanorods. The indices in the diffractogram correspond to the tetragonal structure (COD 1513252) of BaTiO$_3$. Inset **(b)** shows the lattice distortion of the BaTiO$_3$ nanorods, where the dotted lines show the theoretical positions of (200) and (002) reflections in tetragonal BaTiO$_3$. **(c)** Gaussian fits

to the (200) and (002) reflections were performed using Origin software (red lines). Dotted vertical lines indicate the d-spacings determined using X-peak software; the corresponding values are listed above.

The Williamson-Hall method was used to estimate the crystallite size and strain in different crystal directions (see **Fig. 3(a)**). The crystallite size calculated from the Williamson-Hall plot is consistent with values calculated using the Scherrer equation (see **Table I**). The lowest position of the solid lines connecting the positions of (100) and (200) peaks on the Williamson-Hall graph corresponds to a crystallite stretched in the [100] direction as shown in **Fig. 3(a)**. The dimensions of the blocks, perpendicular to the (001) plane in samples S1 and S3, are nearly double compared to the size of blocks perpendicular to the (100) plane, indicating a high anisotropy of these samples. The block sizes in [001] and [100] directions (as well as the block sizes in [002] and [200] directions) of sample S4 are closer in value than in other samples (compare the distance between black circles and black triangles in **Fig. 3(a)** with e.g., the distance between green (red or blue) circles and green (red or blue) triangles therein). The solid line connecting the positions of (100) and (200) peaks in the Williamson-Hall graph is significantly higher for sample S4 than for samples S1, S3, and S5 (the line connecting the two black circles is significantly higher than the lines connecting the circles of the other color in **Fig. 3(a)**). At the same time sample S5, which has the same particle width (50 nm) as the sample S4, reveals the lowest $c/a$ and the smallest lattice strain $e_{001}$ in the [001] direction (see **Table I** and compare the distances between the dotted vertical lines in **Fig. 2(c)**). Thus, we can conclude that the degree of crystal anisotropy is determined not only by the lattice distortion, but also by the other factors related to crystallite shape, their size distribution, and possible curvature, as well the differences in the sample preparation as described in **Section II.A.** Namely, the $Ba^{2+}$ concentration for samples S4 and S5 was 0.3 M and 0.2 M, respectively; and sample S5 was obtained at a Na/Ba ratio of 10 instead of 33.

Consistent with the anisotropic nature of these materials, lattice strain in the [001] direction is 2-2.5 times greater than in the [100] direction for all samples. The higher deformation of the lattice in the [001] direction of samples S1 and S4 compared to S3 and S5 is due to the increased $Ba^{2+}$ concentration in the crystalline particles formed during hydrothermal treatment. However, this difference in lattice strain does not affect the degree of anisotropy and tetragonality of the crystal. Furthermore, samples S1 and S4 have comparable lattice strain values despite significant differences in particle sizes and coherent scattering regions, suggesting that Laplace pressure is not a primary factor determining the tetragonality observed in the smaller particles (e.g., in samples S4 and S5). However, a trend of decreasing tetragonality with decreasing lattice strain in the *c*-direction is observed (see **Fig. 3(b)**), which is consistent with the reported effect of $OH^-$ ions on tetragonality [23]. This trend means that the lattice strain of particles with a diameter greater than 50 nm depends mostly on $OH^-$ ions content rather than on the size effects in the particles.

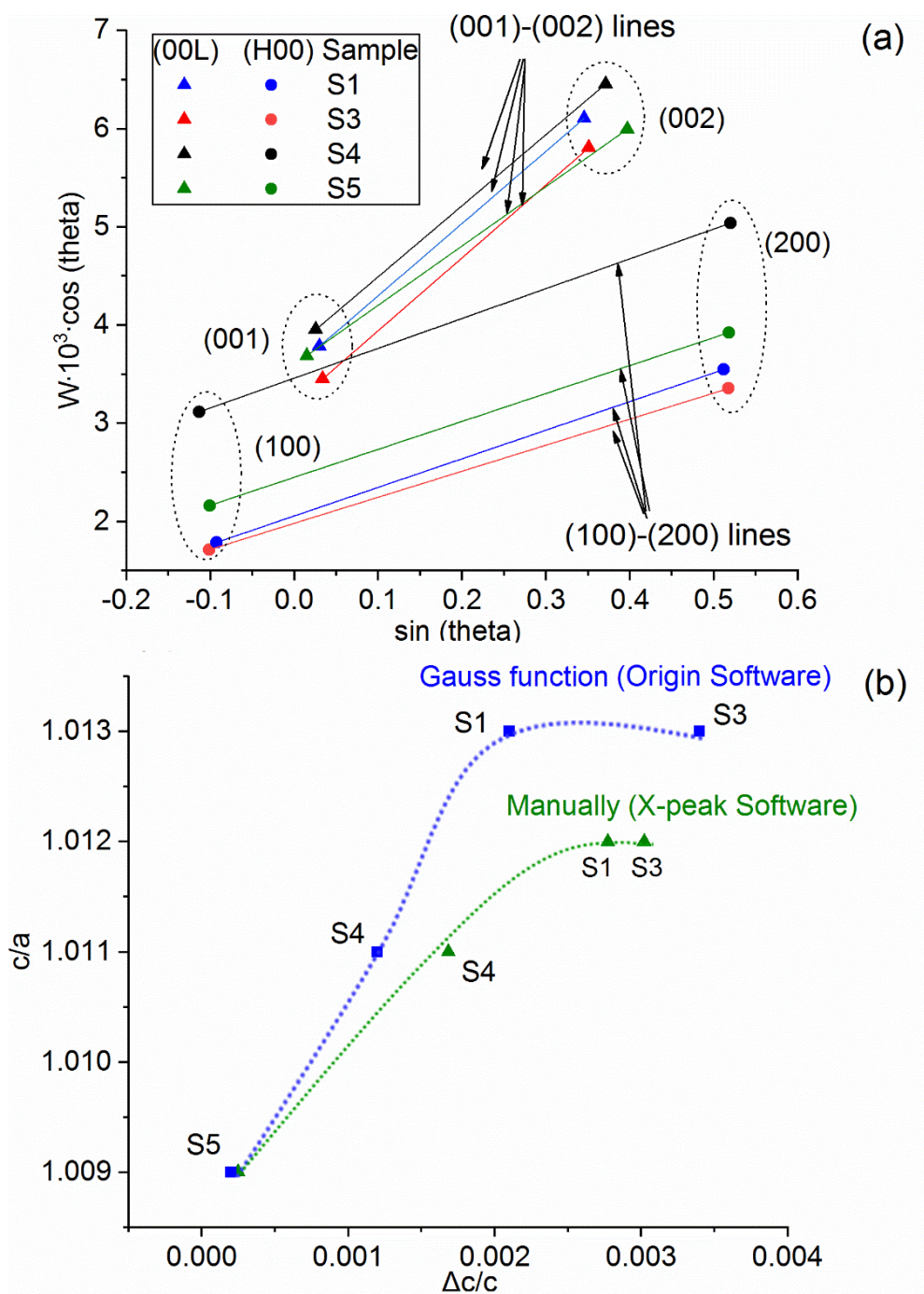

**FIGURE 3**. (a) Williamson-Hall plots for BaTiO$_3$ nanorods (theta is the Bragg angle, W is the diffraction line broadening). (b) The dependence of *c/a* on the lattice distortion calculated using a Gaussian fit and X-peak software.

Sample S5 exhibits lower tetragonality than samples S1, S3, and S4. This is attributed to a slower nucleation rate at lower Na/Ba ratios, likely related to a lower degree of crystallinity (see **Fig. 2**). Additionally, the sample S5 has a crystallite size approximately twice that of sample S4, despite their similar particle sizes (see **Table I**). The narrower particles of the samples S4 and S5 with a narrower particle width distribution (**Fig. 1** and **Table I**) are associated with the increased crystallization rate due to the use of more reactive $TiOCl_2$ instead of $TiO_2$. Sample S4, synthesized with $TiOCl_2$, exhibits longer particles but lower crystal anisotropy compared to sample S1 (synthesized with $TiO_2$). This suggests that even a minimal degree of anisotropy is sufficient to induce anisotropic growth, while the kinetics of particle formation (i.e., nucleation rate and crystal growth speed) determines the final particle sizes.

### C. Nuclear Magnetic Resonance Spectroscopy

$^{137}$Ba NMR spectra of $BaTiO_3$ nanorods were acquired at room temperature and are shown in **Fig. 4**. $^{137}$Ba has a spin of 3/2 and poses a fairly large quadrupole moment $Q = 24.5$ fm$^2$. Nuclear magnetic resonance on quadrupole nuclei can be used to determine the symmetry of the local environment and phase composition of the sample under study. Due to the ionic nature of the chemical bond formed by barium in the studied compound, the shape of the $^{137}$Ba NMR spectra is largely determined by the interaction of its quadrupole moment with the electric field gradient (EFG) created by external charges. Only the 1/2 ↔ -1/2 central transition perturbed by the second-order quadrupole interaction of the $^{137}$Ba nuclei with the EFG is observed. The shift of the central transition, arising solely from the quadrupole interaction, can be expressed as:

$$\nu_{1/2}^{(2)} = -\frac{\nu_Q^2}{16\nu_L}\left(I(I+1) - \frac{3}{4}\right)f_\eta(\theta, \varphi), \qquad (1)$$

where $\nu_Q = -\frac{3C_Q}{2I(2I-1)}$, $C_Q = \frac{eQV_{zz}}{2\pi\hbar}$ is the quadrupole coupling constant, $V_{zz}$ is the largest principal component, and $\eta = \frac{V_{xx}-V_{yy}}{V_{zz}}$ is the asymmetry parameter of the EFG tensor, which determines the effect of quadrupole interactions on the NMR spectra. The function $f_\eta(\theta, \varphi)$ has a cumbersome form and its actual expression can be found elsewhere [24, 25].

The asymmetry parameter $\eta$ has substantially different values for the different possible phases of $BaTiO_3$. Bulk $BaTiO_3$, as well as ceramic samples, pass through four phases at different temperatures: the cubic non-polar phase Pm$\bar{3}$m for temperatures above 393 K ($\eta = 0$), the tetragonal P4mm phase in the temperature range 278-393 K ($\eta = 0$), the orthorhombic C2mm phase between 193-278 K ($\eta = 0.85$), and the rhombohedral R3m phase below 193 K ($\eta = 0$) [26]. However, surface stresses and depolarization fields can shift these temperatures substantially in the nanoparticles due to the large surface/volume ratio. Experimental determination of the asymmetry parameter $\eta$ of the EFG tensor allows, in principle, the determination of the sample's phase composition from the shape of the NMR spectra.

The experimentally observed NMR spectrum is broadened by a distribution of the EFG values. This distribution likely arises from two main sources: (1) elastic deformations caused by oxygen vacancies in the surface layers and OH$^-$ ions into the crystals, and (2) variations in barium ion displacements within domain walls, where the electric polarization changes its value and/or orientation. The change in polarization leads to different displacements of barium ions in the lattice and, accordingly, to different values of the EFG experienced by the barium nucleus.

Since both Ba and Ti are sensitive to local distortions, and the presence of OH$^-$ ions is assumed to be responsible for inducing tetragonality (as noted above), it is therefore plausible that these OH$^-$ ions could also cause a breaking of the local symmetry around these ions. This may lead to a distribution of quadrupolar interactions like those observed in the presence of oxygen vacancies. The observed asymmetric line shape, with a steeper slope on the low-field side and a more gradual decrease on the high-field side, could be attributed to local distortions around the Ba sites induced by OH$^-$ ions. This effect is similar to the distortion caused by the distribution of Sr atoms in Ba$_x$Sr$_{1-x}$TiO$_3$, as reported in Ref. [27].

This suggestion is consistent with several observations. First, the BaTiO$_3$ nanorods exhibit tetragonal distortions (**Fig. 2**). Second, domain structures emerge in sample S3 (as shown by TEM images in Ref. [22]). Third, the observed strain is tensile, whereas oxygen vacancies typically lead to compressive strain. These observations collectively support the hypothesis that OH$^-$ ions, rather than oxygen vacancies, are the primary source of the observed distortions. These distortions lead to the formation of disordered areas within the sample without any specific symmetry inherent to its crystalline structure. However, these disordered areas coexist with "mainly ordered" regions, where the EFG distribution exists, but with a defined symmetry. The obtained spectrum (fit with the red curve) can be decomposed into two spectral lines corresponding to these "distinctly disordered" and "mainly ordered" regions. These lines are labeled as "GIM" (the blue dash-dotted curve) and "Tetr" (the black dash-dotted curve), respectively. The description of abbreviations "GIM" and "Tetr", and a physical nature of the lines are given below.

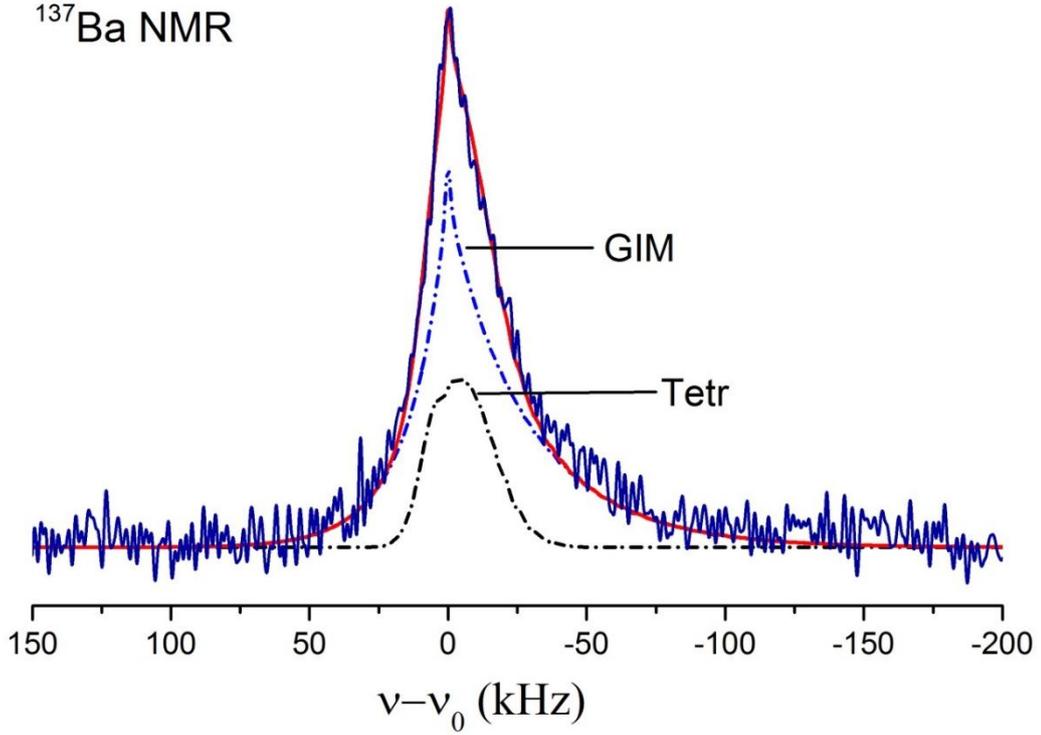

**FIGURE 4.** Fitting of the $^{137}$Ba NMR spectra of BaTiO$_3$ nanorods. Blue solid curves are the measured spectra, and the red solid curve is the fit to the spectra, which is decomposed in two spectral lines labeled as "GIM" (the dash-dotted blue curve) and "Tetr" (the dash-dotted black curve). The "GIM" line represents the Czjzek model, and the "Tetr" line corresponds to the BaTiO$_3$ in the tetragonal phase.

To model the NMR signal from regions where a significant disorder is present, it is necessary to consider the distribution of the quadrupolar parameters $C_Q$ and $\eta$ defined by the principal components of the EFG tensor. Being a traceless symmetric second-rank tensor, EFG has five independent components whose variation due to a random atomic displacement can be described by a Gaussian distribution. This results in the interdependent distribution of $C_Q$ and $\eta$, which define the shape of the NMR spectral line.

The mean values of the EFG tensor components are equal to zero for "distinctly disordered" statistically isotropic areas. In this case, the probability $P(V_{zz}, \eta)$ of the mutual distribution of $C_Q$ and $\eta$ can be obtained using the Czjzek bivariate distribution, also called as the Gaussian isotropic model (GIM), is given by expression [28]:

$$P(V_{zz},\eta) = \frac{V_{zz}^4 \eta}{\sqrt{2\pi}\sigma^5}\left(1-\frac{\eta^2}{9}\right) exp\left[-\frac{V_{zz}^2\left(1+\frac{\eta^2}{3}\right)}{2\sigma^2}\right], \qquad (2)$$

where σ is a single fitting parameter, which represents the width of the Gaussian distribution of the EFG components. This model allows us to calculate the joint probability density function (PDF) of the quadrupolar parameters $V_{zz}$ and $\eta$ used in Eq.(1). The spectral line corresponding to the GIM model (shown by the dash-dotted blue curve in **Fig. 4**) was obtained by summing up the contributions of the individual lines, described by Eq.(1), where the parameter $C_Q$ varies from 0 to 9000 kHz, and the parameter η varies from 0 to 1. The contribution of each individual line into the GIM line was calculated by weighting the intensity of individual lines with the corresponding PDF probability values from Eq.(2). The line corresponding to the signal from disordered regions was chosen to fit the distant edges of the NMR spectrum, where a significant scattering of quadrupolar parameters is present.

The NMR spectral line "Tetr" originated from the "mainly ordered" regions was calculated using the method described in Refs. [29, 30]. In this case the mean values of the principal components of the EFG tensor are not equal to zero and correspond to the symmetry of the crystal field. This method involves applying a Gaussian distribution only to the principal components of the EFG tensor in such a way as to preserve the traceless character of the perturbed EFG. For each set of the principal components, new values of $C_Q$ and $\eta$ were calculated, and the individual lines obtained using Eq.(1) were summed with weights corresponding to the Gauss distribution of the principal components. The best fit to the experimentally measured spectrum was obtained using the EFG with the mean values of principal components corresponding to $C_Q$ = 3.1 MHz and $\eta = 0$. The value $\eta = 0$ corresponds to the tetragonal symmetry, however, the obtained value of the quadrupole constant appeared somewhat higher than its values obtained earlier for BaTiO$_3$ ceramics. For example, the value of $C_Q$ in submicron-sized crystals with tetragonal symmetry is known to increase from 2.22 MHz (at 350 K) to 2.85 MHz (at 295 K) [31], which is accompanied by an increase in the $c/a$ ratio from 1.00996 to 1.0110 [32]. At the same time, the subsequent transition to the orthorhombic phase with a further decrease in temperature leads to a decrease in the quadrupole constant to 2.3 MHz at 250 K. In comparison, the larger quadrupole coupling constant $C_Q$ obtained in our sample indicates a substantial increase in the $c/a$ ratio and suggests a greater degree of tetragonal distortion in the studied nanoparticles.

### III. THEORETICAL MODELING
#### A. The Problem Formulation and Analytical Calculations

The theoretical model presented here focuses on a long BaTiO$_3$ core-shell nanorod with a rectangular cross-section, as shown in **Fig. 5**. The nanorod length $L_s$ is at least 5-10 times longer than its lateral sizes $A_s$ and $B_s$. This shape is consistent with the typical morphology of BaTiO$_3$ nanocrystals and facilitates the construction of a suitable rectangular fine mesh for the FEM. SEM and TEM analysis have confirmed that BaTiO$_3$ nanocrystals commonly exhibit rectangular or near-rectangular morphologies. This is consistent with studies of sintered BaTiO$_3$ structures (e.g., nanocubes, nanorods, and nanoribbons) reported in Refs. [18, 22], **Appendix C** and Chapter 4 in Ref. [33].

The core of the nanorod is modeled as a crystalline material with tetragonal symmetry and is a nearly perfect insulator. The core is covered with a crystalline shell of cubic symmetry, with an average thickness is $\Delta R$. The shell is semiconducting due to the high concentration of free charges and strained due to the high concentration of elastic defects and/or core-shell lattice constant mismatch. The charge screening is formed spontaneously due to multiple mechanisms of spontaneous polarization screening involving both internal and external charges in nanoscale ferroelectrics (e.g., Ref. [34] and refs. therein). The free charges provide effective screening of the core spontaneous polarization and prevent domain formation. The effective screening length in the shell, λ, is very small (less than 1 nm). All numerical calculations in this work are performed at room temperature, whereas the analytical expressions are valid over a wide temperature range.

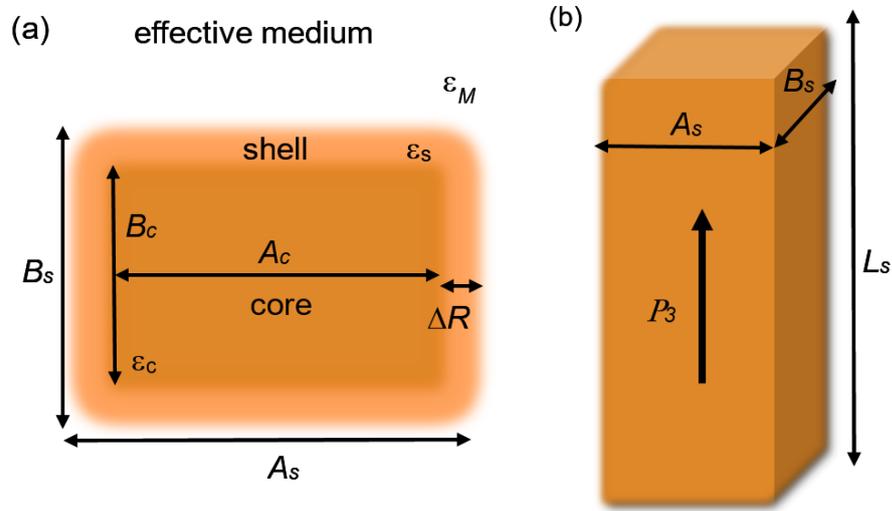

**FIGURE 5.** The cross-section (**a**) and the side view (**b**) of the BaTiO$_3$ nanorod. The crystalline core permittivity is $\hat{\varepsilon}_c$. The core is covered with a crystalline shell of cubic symmetry, with permittivity $\varepsilon_s$ and average thickness $\Delta R$.

Analytical expressions are used to calculate the elastic stresses and strains in the long core-shell nanorod, as derived in Ref. [35]. These expressions are highly accurate for rods with a cylindrical cross-section. For rods with an arbitrary cross-section, the accuracy is only slightly reduced, provided the aspect ratio (length divided by width) is greater than 10. For a very long nanorod or nanowire, the variation $\Delta c$ of the lattice constant $c$, the variation $\Delta a$ of the lattice constant $a$, and the tetragonality ratio of the lattice constants $c$ and $a$, are given by expressions [35]:

$$\frac{\Delta c}{c} \approx u_{33} \approx u_c + (1 - \delta V)Q_{11}P_3^2 + \delta V \delta u, \tag{3a}$$

$$\frac{\Delta a}{a} \approx u_{11} \approx u_c + (1 - \delta V)Q_{12}P_3^2 + \delta V \left[\delta u + \frac{-(s_{11}-s_{12})\delta u + (s_{11}Q_{12} - s_{12}Q_{11})P_3^2}{2(s_{11}+s_{12})}\right], \tag{3b}$$

$$\frac{c}{a} \approx \frac{1+u_{33}}{1+u_{11}} \approx 1 + (1-\delta V)(Q_{11}-Q_{12})P_3^2 - \frac{1}{2}\delta V\left[\frac{s_{11}Q_{12}-s_{12}Q_{11}}{s_{11}+s_{12}}P_3^2 - \frac{s_{11}-s_{12}}{s_{11}+s_{12}}\delta u\right]. \quad (3c)$$

Here, $u_{ij}$ are the components of elastic strain tensor, $u_c$ is the core strain, $P_3$ is the axial spontaneous polarization of the rod, $Q_{ij}$ are electrostriction coefficients, and $s_{ij}$ are elastic compliances. Also, the relative shell volume ($\delta V$) and "effective" strain ($\delta u$) are introduced as:

$$\delta V = \frac{V_s}{V}, \qquad \delta u = u_s - u_c. \quad (4)$$

In Eq.(4), the shell volume is $V_s$ and the nanorod volume is $V$. In a general case, the effective strain $\delta u$ is created not only by the difference between the core and the shell chemical strains ($u_c$ and $u_s$), as assumed in Eq.(4), but also by the lattice constants mismatch and/or different thermal expansion coefficients in the core and the shell. Due to the reasons discussed in Ref. [35], the surface tension contribution to $\delta u$ and $P_3^2$ appear negligibly small (less than 0.01 %) for the considered transverse sizes of the nanorods (50 nm and more, see **Table I**) and realistic values of the surface tension coefficient $\mu < 4$ N/m. The first two terms in Eq.(3) coincide with corresponding expressions for a bulk ferroelectric with the spontaneous polarization $P_3$ in the tetragonal ferroelectric phase. Next terms, proportional to the relative shell volume $\delta V$, are caused by the elastic anisotropy between the tetragonal core and cubic shell, as well as by the effective chemical strain, $\delta u$. From Eq.(5), the non-zero $\frac{\Delta c}{c}$ and $\frac{c}{a}$ can exist in paraelectric core-shell nanorods due to the term $\delta u\, \delta V$. The polarization $P_3$ is described by the time-dependent Landau-Ginzburg-Devonshire (LGD) equation, which must be solved together with the Poisson equation for depolarization electric field and the equations of state for elastic strains (see details in Ref. [35]).

The tetragonality $\frac{c}{a}$ and the ratio $\frac{\Delta c}{c}$ as a function of relative shell volume $\delta V$ and effective strain $\delta u$ is shown in **Fig. 6(a)** and **6(b)**, respectively. The color maps are calculated for long BaTiO$_3$ nanorods at room temperature. Tensile strains support and/or induce the ferroelectric (FE) state in the rods (see the part of the maps where $\frac{c}{a} > 1$ and $\frac{\Delta c}{c} > 0$); compressive strains suppress the FE state and/or induce the paraelectric (PE) state in the rods (see the part of the maps, where $\frac{c}{a} < 1$ and $\frac{\Delta c}{c} < 0$). The nature of the FE-PE transition depends on the type of strain. Compressive strains lead to first order transitions, as evidenced by the relatively sharp boundary in **Fig. 6(a)** and **6(b)**. In contrast, tensile strains result in second order transitions, characterized by the diffuse boundary in **Fig. 6(a)** and **6(b)**.

The order of the FE-PE transition changes at the critical point, $\{\delta u_{cr}, \delta V_{cr}\}$, where the "renormalized" expansion coefficient $\beta_R$ in the effective Landau energy becomes zero. The effective Landau energy is given by:

$$F = \frac{\alpha_R}{2}P_3^2 + \frac{\beta_R}{4}P_3^4 + \frac{\gamma}{6}P_3^6 + \frac{\delta}{8}P_3^8 - E_3^e P_3, \quad (5)$$

where $\alpha_R \sim 2\left\{a_1 - \delta u\, \delta V \frac{Q_{11}+Q_{12}}{s_{11}+s_{12}}\right\}$ and $\beta_R = 4\left\{a_{11} + \delta V \frac{s_{11}(Q_{11}^2+Q_{12}^2)-2s_{12}Q_{11}Q_{12}}{2(s_{11}^2-s_{12}^2)}\right\} - 8\frac{Z_{211}}{s_{11}+s_{12}}\delta V \delta u$ (see details in Ref. [35]).

Despite $\alpha_R$ and $\beta_R$ depending linearly on $\delta V$, the FE-PE transition order does not appear to depend on the relative shell volume $\delta V$ as demonstrated in **Fig. 6(a)** and **6(b)**. Namely, the nanorod core is ferroelectric, but the FE-PE transition is absent for the small values $0 \leq \delta V < 0.2$. This absence is due to the negative values of $\alpha_R$ and $\beta_R$ across the strain range ($-2\,\% < \delta u < 2\,\%$), resulting in $\alpha_{11} < 0$ and $a_{11} < 0$. For intermediate values of $\delta V$ ($0.2 < \delta V < 0.7$) and under compressive strains, the first-order FE-PE transition appears in the nanorod core (because $\beta_R$ changes sign in the range of $\delta V$ and $\delta u$). For large values of $\delta V$ ($\delta V > 0.7$) and under compressive strains, the second-order FE-PE transition appears in the nanorod core (because $\beta_R > 0$ in this range of $\delta V$ and $\delta u$). The behavior of the transition order with increasing $\delta V$ (i.e., no transition → first order transition → second order transition) determines the step-like dependencies of the tetragonality and lattice strains on $\delta u$ in the intermediate range of $\delta V$ (i.e., for $0.2 < \delta V < 0.7$).

Dependencies of $\frac{c}{a}$ and $\frac{\Delta c}{c}$ on the effective strain $\delta u$ calculated for different relative shell volumes $\delta V$ at room temperature, are shown in **Fig. 6(c)** and **6(d)**, respectively. The dependencies calculated for small ($\delta V = 0.1$) and large ($\delta V = 0.9$) relative shell volumes are quasi-linear (see red and green curves in **Fig. 6(c)** and **6(d)**), while the dependencies calculated for $0.2 < \delta V < 0.7$ contain a step-like feature corresponding to the first order PE-FE phase transition (see blue and magenta curves in **Fig. 6(c)** and **6(d)**). The slope of $c/a(\delta u)$ and $\Delta c/c(\delta u)$ curves increases with an increase in $\delta V$; the curves intersect at $\delta u \approx 1\,\%$, and $\delta u \approx 0.78\,\%$, respectively. The intersections correspond to the vertical straight contour lines shown by thin dotted lines in **Fig. 6(a)** and **6(b)**. The vertical lines correspond to the "saddle" contour lines, where the derivatives of the functions $c/a$ and $\Delta c/c$ with respect to the relative shell volume $\delta V$ are zero. The derivatives $\frac{\partial (c/a)}{\partial (\delta V)}$ and $\frac{\partial (\Delta c/c)}{\partial (\delta V)}$ are negative to the left of the saddle contour lines and positive to the right of the lines. Zero derivatives indicate that the values of $c/a$ and $\Delta c/c$ are $\delta V$-independent along the vertical lines, which corresponds to the intersection of the curves in **Fig. 6(c)** and **6(d).**

The quasi-linear dependencies of $\frac{c}{a}$ and $\frac{\Delta c}{c}$ on strain, at small values of $\delta V$ ($0 \leq \delta V < 0.2$), exhibit a small slope, which corresponds to the absence of the FE-PE transition in the ferroelectric core. The quasi-linear dependencies of $\frac{c}{a}$ and $\frac{\Delta c}{c}$ on strain, at large values of $\delta V$ ($\delta V > 0.7$), exhibit a large slope, which corresponds to a possible second-order FE-PE transition in the core. The step-like dependencies of $\frac{c}{a}$ and $\frac{\Delta c}{c}$ on strain, at intermediate values of $\delta V$ ($0.2 < \delta V < 0.7$), correspond to the first-order FE-PE phase transition in the core (the step itself corresponds to the point where the transition order changes).

The slope of $\frac{c}{a}$ and $\frac{\Delta c}{c}$ vs. strain curves increases monotonically with increasing $\delta V$. This is because the slope is mainly determined by the coefficient $\alpha_R$ (rather than by the coefficient $\beta_R$), which depends on $\delta V$ according to $\alpha_R \sim 2\left\{a_1 - \delta u\, \delta V \frac{Q_{11}+Q_{12}}{s_{11}+s_{12}}\right\}$. As $\delta V$ increases, the dependence of $\alpha_R$ on the stain becomes stronger. This trend contrasts with the above behavior of the transition order (determined by the sign of $\beta_R$), which changes from "no transition" to a "first order transition" and then to a "second order transition" as the shell volume increases.

Let us underline that the step-like behavior of the $\frac{c}{a}$ and $\frac{\Delta c}{c}$ curves in **Fig. 6(c)** and **6(d)** aligns well with the critical point where $\beta_R$ vanishes. A sharp jump in $\frac{c}{a}$ and $\frac{\Delta c}{c}$ curves can be seen where Landau theory predicts a change of the PE-FE transition order. The experimental verification of the prediction can validate the proposed model. In a favorable case, the transition order in the BaTiO$_3$ nanorods can be controlled by changing the strain or/and the shell volume. The strain control of the transition order is an effective way to tune the nanorods' dielectric properties for various applications.

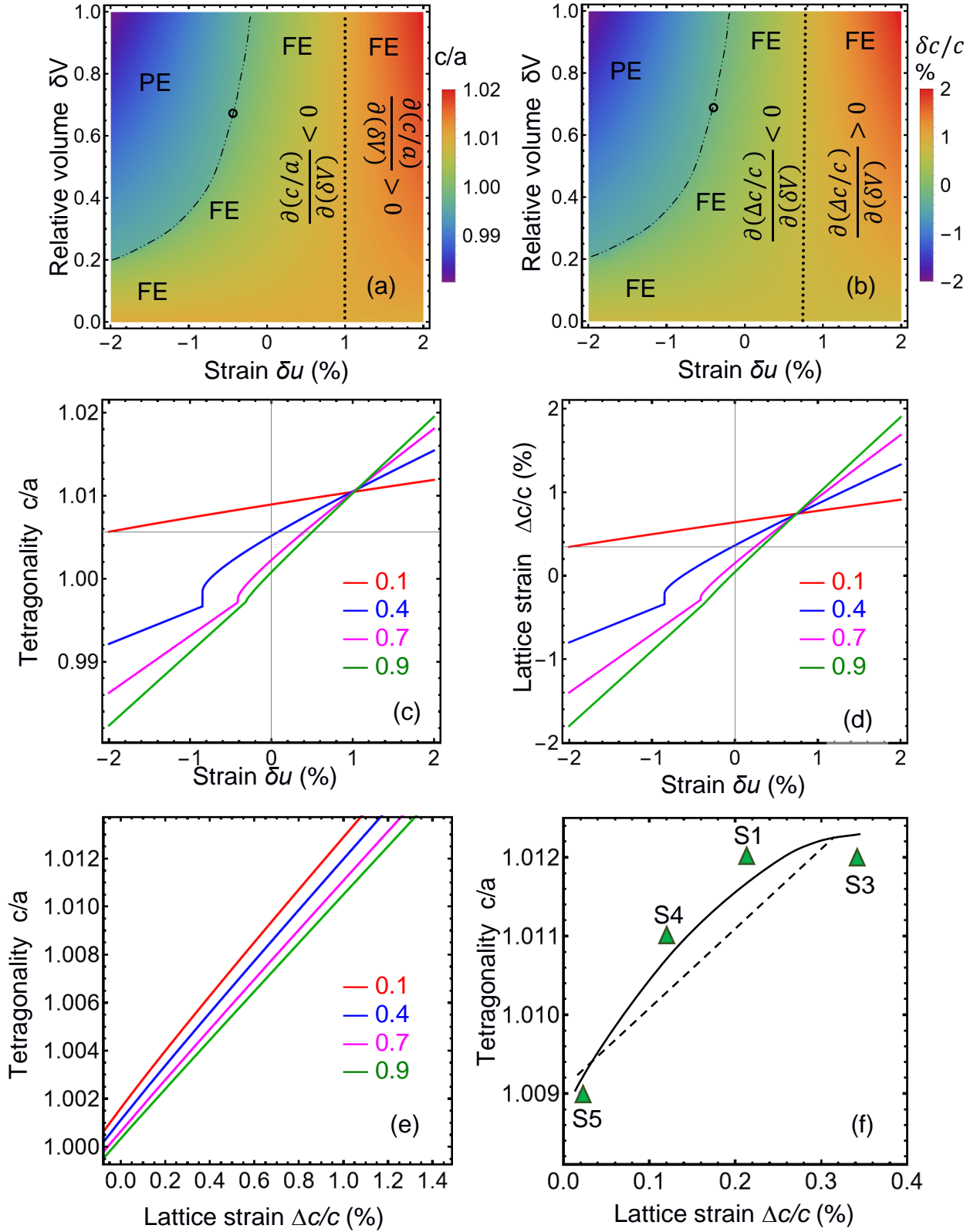

**FIGURE 6.** The tetragonality $\frac{c}{a}$ **(a)** and the lattice strain $\frac{\Delta c}{c}$ **(b)** as a function of relative shell volume $\delta V$ and effective strain $\delta u$ calculated in long BaTiO$_3$ nanorods at room temperature $T = 298$ K and $u_c = 0$. Color scales are $\frac{c}{a}$ and $\frac{\Delta c}{c}$ values in dimensionless units and %, respectively. The FE-PE transition occurs at the dash-dotted curve. The critical point, where the FE-PE transition changes its order, is shown by the black open circle. The bottom part of the curve (below the critical point), where the polarization contrast changes from yellow-green to green-blue more sharply, corresponds to the first-order FE-PE transition; the top part of the curve (above the critical point), where the polarization contrast changes from yellow-green to green-blue continuously, corresponds to the second-order FE-PE transition. The tetragonality $\frac{c}{a}$ (c) and the lattice strain $\frac{\Delta c}{c}$ (d) as a function of the effective strain $\delta u$. The tetragonality $\frac{c}{a}$ as a function of the lattice strain $\frac{\Delta c}{c}$ (e). Plots (c), (d), and (e) are calculated for $T$ =298 K, $u_c = 0$, and different values of $\delta V = 0.1$ (red curves), 0.4 (blue curves), 0.7 (magenta curves) and 0.9 (green curves). (f) The experimental dependence of $\frac{c}{a}$ versus $\frac{\Delta c}{c}$ (shown by green triangles) was fit by the least squares method. The dashed black line corresponds to a linear fit; the solid black curve corresponds to a parabolic fit (fitting parameters are described in the text).

The curves for $\frac{c}{a}$ in **Fig. 6(e)** are calculated for several values of $\delta V$, with the assumption that $\frac{\Delta c}{c} = (1 - \delta V)Q_{11}P_3^2 + \delta V \delta u$ in accordance with Eq.(3a), where we set $u_c = 0$. From these curves, the dependence of $\frac{c}{a}$ versus $\frac{\Delta c}{c}$ is linear, and the values of $\frac{\Delta c}{c}$ required to overcome the tetragonality of the bulk sample ($\frac{c}{a} \approx 0.011$) must be at least 0.8 % for $\delta V = 0.1$ and 1 % for $\delta V = 0.9$. However, **Fig. 6(f)** shows that $\frac{c}{a} \approx 0.011$ can be reached with much smaller values of $\frac{\Delta c}{c} \approx 0.13$ %, which is consistent with XRD data (green triangles) in **Fig. 3(b)**.

It is worth noting that the condition $u_c = 0$ can be violated easily. Also, Eqs.(3) are derived for the constant $\delta u$ and $\delta V$; the derivation does not consider different orientation of the nanorods in the powder sample, as well as the strain gradient and possible curvature of the long rods (considered in the next subsection). The statistical averaging of Eqs.(3) over all possible orientations of the rods leads to the following expressions for the mean square values, which are registered by XRD:

$$\sqrt{\left\langle\left(\frac{\Delta c}{c}\right)^2\right\rangle} \approx \sqrt{\frac{1}{3}\langle u_{11}^2 + u_{22}^2 + u_{33}^2\rangle} \approx \langle u_c\rangle + (1-\delta V)Q_{11}\langle P_s^2\rangle + \delta V\langle \delta u\rangle, \quad (6a)$$

$$\sqrt{\left\langle\left(\frac{\Delta a}{a}\right)^2\right\rangle} \approx \sqrt{\frac{1}{3}\langle u_{11}^2 + u_{22}^2 + u_{33}^2\rangle} \approx \langle u_c\rangle + (1-\delta V)Q_{12}\langle P_s^2\rangle + \delta V\langle \delta u\rangle +$$
$$\delta V\left[\frac{-(s_{11}-s_{12})\langle\delta u\rangle+(s_{11}Q_{12}-s_{12}Q_{11})\langle P_s^2\rangle}{2(s_{11}+s_{12})}\right] \quad (6b)$$

$$\sqrt{\left\langle\left(\frac{c}{a}\right)^2\right\rangle} \approx \sqrt{\frac{1}{6}\sum_{i\neq j}\left\langle\left(\frac{1+u_{ii}}{1+u_{jj}}\right)^2\right\rangle} \approx 1 + (Q_{11}-Q_{12})(1-\delta V)\langle P_s^2\rangle + \frac{s_{11}-s_{12}}{2(s_{11}+s_{12})}\delta V\langle \delta u\rangle +$$
$$\frac{s_{12}Q_{11}-s_{11}Q_{12}}{2(s_{11}+s_{12})}\delta V\langle P_s^2\rangle. \quad (6c)$$

Here, the symbol $\langle ...\rangle$ denotes statistical averaging, and the quantities $\langle P_s^2\rangle$, $\langle\delta u\rangle$, and $\delta V$ are the fitting parameters corresponding to the averaged values of the square of the spontaneous polarization, effective strain, and relative shell volume. Each of the samples S1, S3, S4, and S5 can be characterized by its own set of these parameters, whose values can be obtained by fitting Eqs.(6) to the data listed in **Table I**.

**Figure 6(f)** shows the least squares fit of the experimental $\frac{c}{a}$ versus $\frac{\Delta c}{c}$ data (green triangles) using Eqs.(6). The dashed black line corresponds to the linear fit, $\frac{c}{a} = C_0 + C_1\left(\frac{\Delta c}{c}\right)$, with fitting parameters $C_0$ and $C_1$. The solid black curve corresponds to the parabolic fit, $\frac{c}{a} = C_0 + C_1\left(\frac{\Delta c}{c}\right) + C_2\left(\frac{\Delta c}{c}\right)^2$, with fitting parameters $C_0$, $C_1$, and $C_2$. Using the fitting curves (shown in **Fig. 6(f)**) and Eqs.(6), we determined the average polarization $\sqrt{\langle P_s^2 \rangle}$ and the shell strain $\langle \delta u_s \rangle$ as a function of the shell relative volume $\delta V$ (see **Fig. B1** in **Appendix B**). The fitting, whose reliability is sufficiently high, yields an average of 46 – 63 μC/cm², which is nearly twice the bulk value of BaTiO$_3$ at room temperature ($P_s \approx$ 24 – 27 μC/cm²), but significantly smaller in comparison with the giant spontaneous polarization (above 100 – 120 μC/cm²) experimentally observed in BaTiO$_3$ nanocubes [18]. The fitted average polarization increases monotonically from 46 – 49 μC/cm² to 59 – 63 μC/cm² with an increase in $\delta V$ from 0.1 to 0.5 for samples S1-S5 (see **Fig. B1(a)**). The polarization enhancement is conditioned by the shell strain, which is compressive and decreases monotonically with increase in $\delta V$ (see **Fig. B1(b)**). The shell strain exceeds very high values 10 – 20 % for $\delta V < 0.2$ and becomes smaller than 3 – 4 % for $\delta V > 0.5$ in all samples S1-S5. At the same time the average strain of the nanorod, $\langle \delta u_s \rangle \delta V + \langle u_c \rangle (1 - \delta V)$, does not exceed 2.3 %, which is a reasonable value (see inset in **Fig. B1(b)**).

Although the fitting results may seem counterintuitive, this can be explained by the fact that the experimental tetragonality ($c/a \approx$ 1.009–1.013) is close to the bulk BaTiO$_3$ value ($c/a \approx$ 1.010), and the average spontaneous polarization $\sqrt{\langle P_s^2 \rangle} \approx$ 46 − 63 μC/cm² is significantly enhanced (consistent with previous observations [18]). However, the experimental spontaneous strain ($\Delta c/c \approx$ 0.0012– 0.0034), which represents the overall strain in the nanorod, is significantly lower than the corresponding bulk value ($\Delta c/c \approx$ 0.0075). This contrasts with the fitting results, which suggest a much larger strain within the shell of the nanorod. Despite this, the fitting results indicate a much higher strain within the nanorod's shell. This suggests that the fitting process, particularly for thin shells ($\delta V <$ 0.2), might be overestimating the shell strain because it assumes a coordinate-independent shell strain and may not fully account for other complex effects, including OH⁻ ions, within the nanorod shell.

## B. Finite Element Modeling of Curvature-Induced Changes of the Strain-Gradient, Polarization Distribution, and Domain Morphology in BaTiO$_3$ Nanorods

FEM simulations are performed using COMSOL Multiphysics® software. The COMSOL model uses the electrostatics module to solve the Poisson equation; solid mechanics and general math (PDE toolbox) modules are used in conjunction to achieve a self-consistent solution for the time-dependent LGD equations. The FEM analysis is conducted using various discretization densities for the self-adaptive rectangular mesh and different initial polarization distributions, such as randomly small fluctuations or poly-domain states. The material parameters used for BaTiO$_3$ in the FEM simulations are detailed in **Table AI** in **Appendix A**. The mesh element size ranges from a minimum of 0.4 nm (equal to the unit cell size of BaTiO$_3$ at room temperature) to a maximum of 0.8 nm. The surrounding dielectric medium is modeled with a larger mesh size of 4 nm. This increase in the mesh size resulted in only minor changes in the electric polarization, electric field, and elastic stress/strain distribution.

FEM simulations are performed for rectangular-shaped nanorods with varying sizes and aspect ratios. The length of the rod $L$ varies from 20 nm to 40 nm. Lateral dimensions ($A$ and $B$) range from 4 nm to 12 nm, resulting in aspect ratios between 3 – 10. The screening length $\lambda$ varies from 0.1 nm to 1 nm. As a rule, the increase of $\lambda$ from 0.1 nm to 1 nm leads to the instability of the single domain state and induces the formation of various domain morphologies [35]. The initial distribution of polarization is a single domain wall with randomly small fluctuations. Final stable structures were obtained after a long simulation time, $t \gg 10^3 \tau$, where the parameter $\tau$ is the Landau-Khalatnikov relaxation time [36]. The nanorod is clamped at one side ($z = 0$), and a shear force with the density $F$ is applied vertically along the y-axis to the opposite side ($z' = L$). The corresponding geometry of the undeformed and curved nanorod is shown in **Fig. 7.**

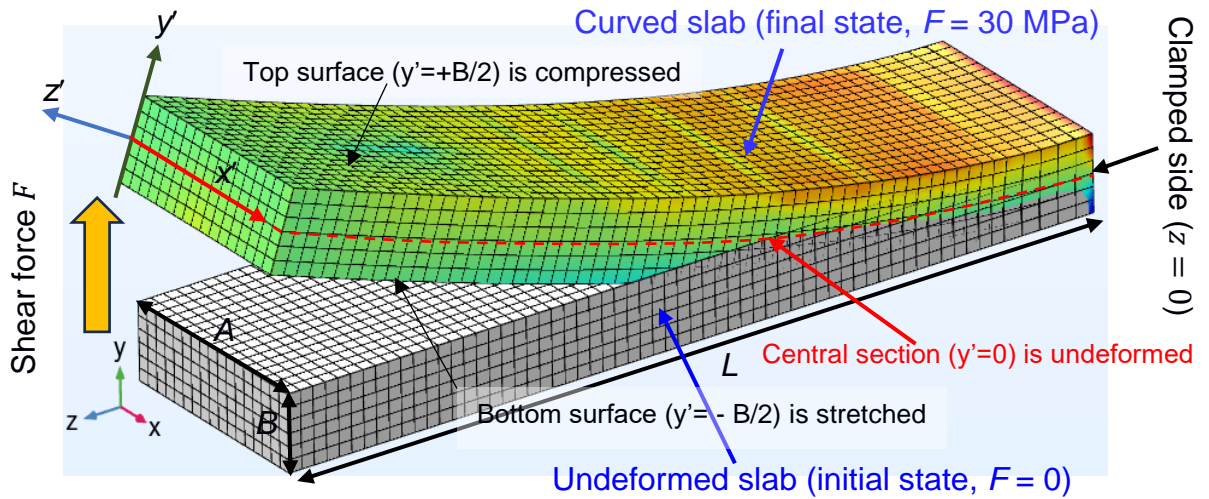

**FIGURE 7.** Geometry of undeformed (grey parallelepiped) and curved (colored parallelepiped) rectangular-shaped nanorod with sizes $B < A < L$, hereinafter referred to as nano-slab. The nano-slab is clamped at the right side ($z = 0$) and the shear force is applied vertically to the left side ($z' = L$). The color scale of the curved nano-slab shows the example of the surface distribution of hydrostatic stress, namely $\sigma_{11} + \sigma_{22} + \sigma_{33}$, calculated when the shear force density $F = 30$ MPa. The figure also shows the reference coordinate frame $\{x, y, z\}$ and the local frame $\{x', y', z'\}$ attached to a specific point on the nano-slab. The bottom surface ($y' = -B/2$) is stretched, the central section ($y' = 0$) is undeformed, and the top surface ($y' = +B/2$) is compressed.

One particularly interesting nanorod shape for applications corresponds to the nano-slabs with dimensions $B < A < L$, whose electric polarization and elastic strain distributions are discussed below. We consider both undeformed (i.e., mechanically free, $F = 0$) slabs (see e.g., **Fig. 8**) and the slabs clamped on one side to a substrate and subsequently deformed/curved by a shear force with the density $F$ applied to the other side (see e.g., **Figs. 9** and **10**). In the latter case, we increase the force density $F$ until domain splitting begins.

The domain structure is determined by the depolarization field in mechanically free slabs, as illustrated in **Figs. 8(a)-(c)**. It is seen from **Fig. 8(a)** that the so-called oppositely polarized closure domains of the polarization component $P_1$ appear near opposite ends of the slab (i.e., near $z = 0$ and $z = L$), where the cross-sectional area $AB$ is smallest. This domain configuration minimizes the depolarization field outside of the slab by allowing the polarization component $P_1$ to "circulate" along the surface (**Fig. 8(a)**). The polarization component $P_2$ (directed along the smallest side of the rod) is absent (**Fig. 8(b)**), and the polarization component $P_3$ (directed along the longest side of the rod) consists of two opposite domains (**Fig. 8(c)**). The elastic strain components are distributed relatively uniform inside the central part of the free-standing nano-slab; inhomogeneous strains are localized inside the closure domains and near the domain walls, consistent with the known effect of stress concentration in these regions (see **Figs. 8(d)**, **8(e)**, and **8(f)**).

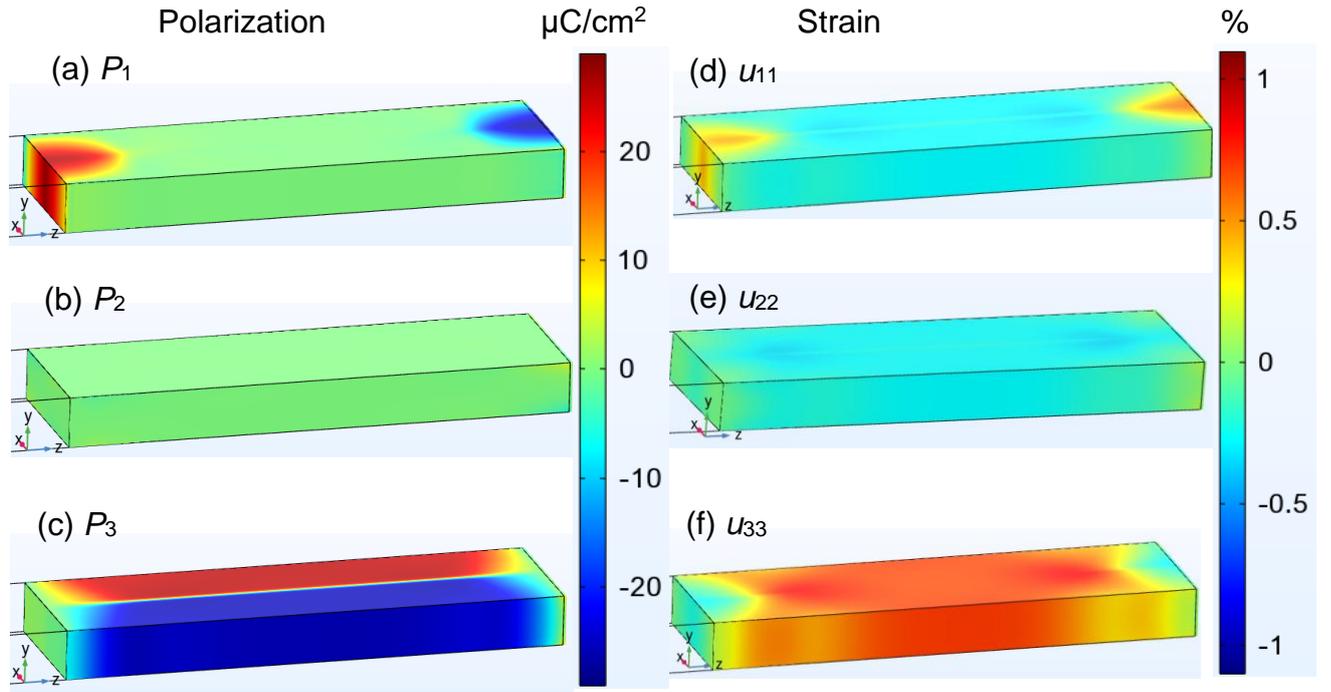

**FIGURE 8**. The distribution of the polarization components **(a)** $P_1$, **(b)** $P_2$, and **(c)** $P_3$, and the strain components **(d)** $u_{11}$, **(e)** $u_{22}$, and **(f)** $u_{33}$ calculated at the surface of a free-standing BaTiO$_3$ nano-slab with dimensions $4 \times 12 \times 40$ nm$^3$ at room temperature. The screening length $\lambda = 1$ nm.

For small curvatures, only the polarization gradient is observed across the nano-slab thickness, without significant changes in the domain structure morphology. However, when the bending force exceeds a critical threshold, an isomorphic phase transition occurs in the domain structure [8]. The transition is driven by the tendency to minimize the slab's elastic energy through domain splitting.

To illustrate the above statement, **Fig. 9** shows the polarization and strain distributions in the nano-slab clamped at the side $z = 0$, a vertical shear force with the density of 30 MPa applied to the opposite side $z' = L$. An elastic stress gradient appears throughout the depth of the curved slab along the bending direction; one side of the slab is stretched, while the other side is compressed (see **Figs. 9(d)**, **9(e),** and **9(f)**). The 2D-surface with zero strain corresponds to the so-called "central section", which is close to the cross-section $y' = 0$ in the local coordinate frame $\{x', y', z'\}$ (shown by the red dashed lines in **Fig. 7**). Electrostriction coupling causes significant changes in the domain structure with an increase of the strain gradient. Stretching the nano-slab along the polarization direction increases its value, while compression decreases it. Considering that the strains in opposite directions have opposite signs (due to the positive Poisson coefficient in BaTiO$_3$), domains with polarization perpendicular to the original direction occur in the compressed half of the nano-slab and become pronounced with an increase in bending degree (see **Figs. 9(a)**, **9(b),** and **9(c)**).

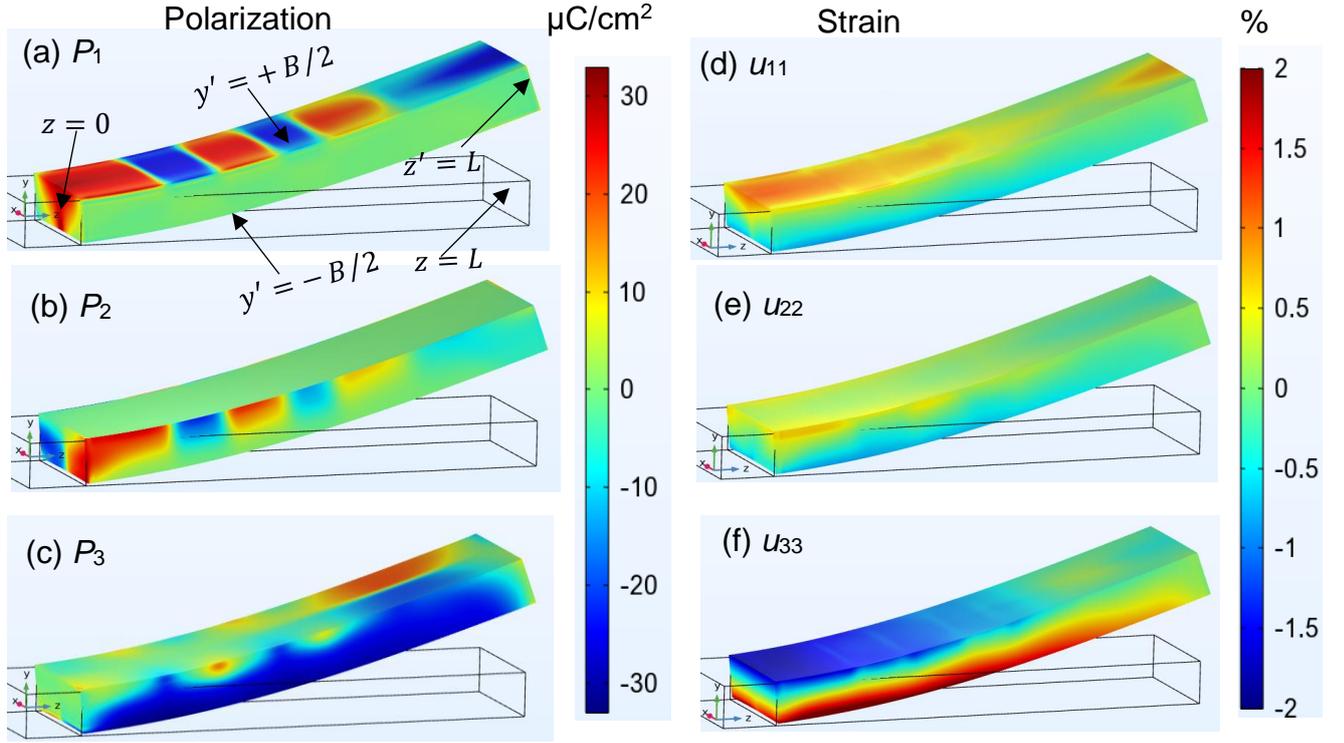

**FIGURE 9**. The distribution of the polarization components **(a)** $P_1$, **(b)** $P_2$, and **(c)** $P_3$, and the strain components **(d)** $u_{11}$, **(e)** $u_{22}$, and **(f)** $u_{33}$ calculated at the surface of a curved BaTiO$_3$ nano-slab with dimensions $4 \times 12 \times 40$ nm$^3$ at room temperature and for a screening length $\lambda = 1$ nm. The slab is clamped at the side $z = 0$ and the shear force with the density of 30 MPa is applied vertically to the side $z' = L$. The empty rectangular-shaped parallelepiped shows the initial state of the undeformed slab ($F = 0$).

The distribution of the polarization components $P_i$ and diagonal strains $u_{ii}$ calculated for the different cross-sections of the BaTiO$_3$ nano-slab curved by a shear force $F$ with the density of 30 MPa, are shown in **Fig. 10**. Each group of three images (from the left to the right) corresponds to: (1) the maximally stretched bottom cross-section ($y' = -B/2$), (2) the less stretched middle cross-section ($y' = +B/4$) and (3) the maximally compressed top cross-section ($y' = +B/2$). These cross-sections are curved being perpendicular to the smallest side of the slab $\{A, B\}$ in the local frame $\{x', y', z'\}$ (shown in **Fig. 7**). The domain stripes are stable in the compressed part of the slab (at $y' > 0$), and they are absent in the stretched part (at $y' < 0$).

The distribution of $P_1$ (shown in **Fig. 10(a)**) correlates with the distribution of $u_{11}$, as evident in the reproduction of the closure domains and domain stripes in **Fig. 10(b)**. A vertical central line, visible in the $u_{11}$ cross-section at $y' = -B/2$, corresponds to the domain wall between $P_3$ domains, as seen in **Fig. 10(e)**. Distributions of $P_2$ and $u_{22}$ have very small contrast and look uncorrelated (compare **Fig. 10(c)** and **10(d)**). Distributions of $P_3$ contain two oppositely polarized domains separated by the wall, which are pronounced above the central line (i.e., at $y' < 0$). The $P_3$ domains gradually disappear near the clamped side with an increase in $y'$ (see **Fig. 10(e)**). The distribution of $P_3$ very weakly correlates with $u_{33}$ distribution (compare **Fig. 10(e)** and **10(f)**). However, a thin lighter central line, which corresponds to the $P_3$-domain wall, is seen in the distributions of $u_{11}$ and $u_{22}$ calculated for $y' = -B/2$, which are shown in **Fig. 10(b)** and **Fig. 10(d)**, respectively.

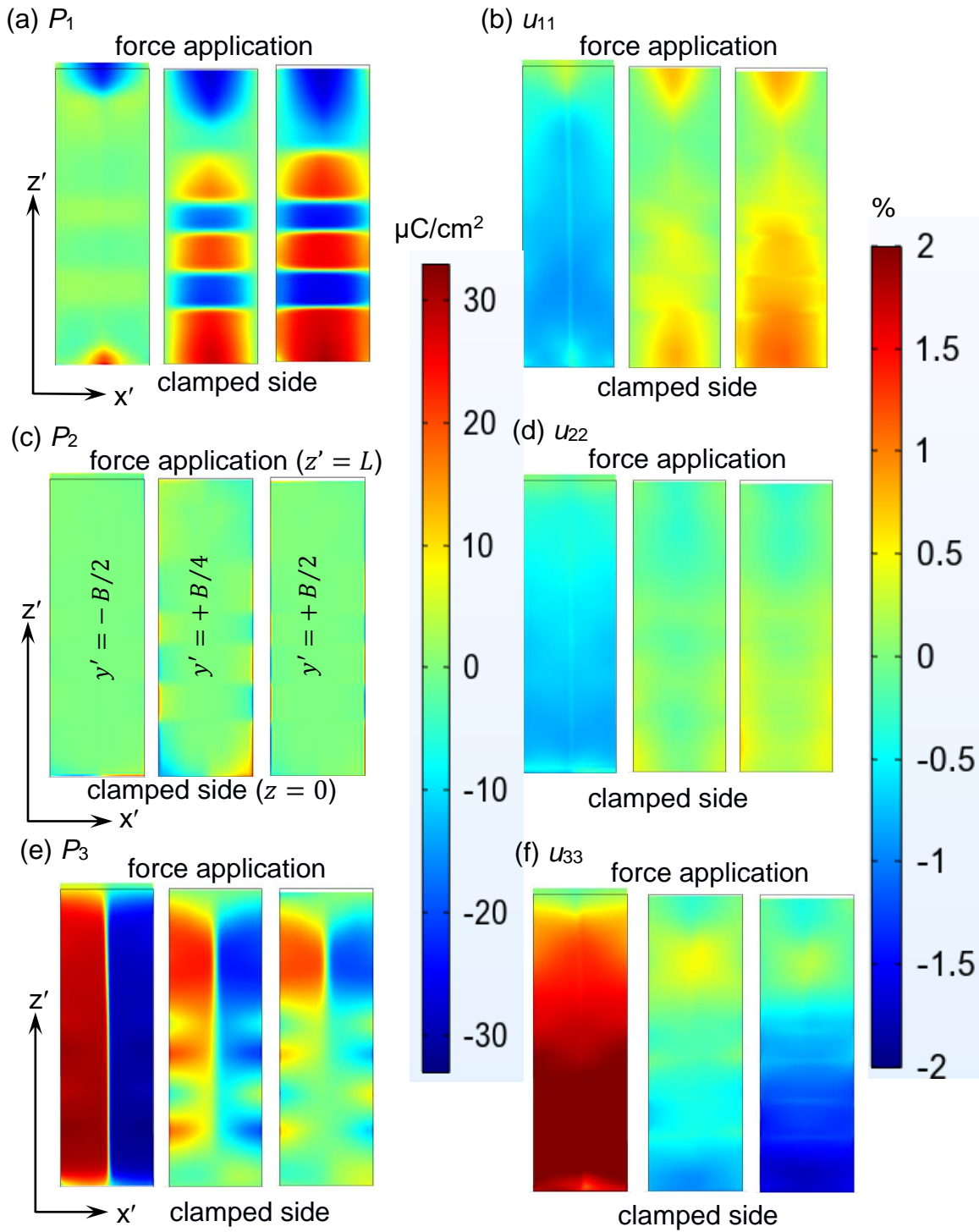

**FIGURE 10**. The distribution of the polarization components **(a)** $P_1$, **(c)** $P_2$, and **(e)** $P_3$, and the strain

components **(b)** $u_{11}$, **(d)** $u_{22}$, and **(f)** $u_{33}$ **(c)** calculated in the different cross-sections of a curved BaTiO$_3$ nano-slab with dimensions $4 \times 12 \times 40$ nm³ at room temperature and for $\lambda = 1$ nm. Each group of three images presents cross-sections of the slab at different depths: $y' = -B/2$ (stretched bottom surface), $y' = +B/4$ (very slightly compressed middle section), and $y' = +B/2$ (compressed top surface), taken perpendicular to the slab's smallest dimension $y'$. The slab is clamped at one side ($z' = 0$) and a shear force $F$ with the density of 30 MPa is applied vertically to the opposite side ($z' = L$). The empty rectangular grey boxes show the initial cross-sections of the undeformed slab.

The distributions of the polarization components $P_1$ and $P_3$, calculated for different cross-sections of the curved BaTiO$_3$ nano-slab, are shown in the left and right columns of **Fig. 11**, respectively. The slab is clamped at the side $z = 0$ and a shear force is applied to the opposite side $z' = L$. The force density $F$ varies from 3 to 60 MPa. Each group of images consists of five $\{x', z'\}$ cross-sections perpendicular to the smallest dimension of the slab at different depths $y'$, representing varying degrees of bending: $y' = -B/2$ (the bottom surface that is maximally stretched), $y' = -B/4$ (the middle cross-section, which is less stretched), $y' = 0$ (the central cross-section, which is undeformed), $y' = +B/4$ (the middle cross-section, which is compressed), and $y' = +B/2$ (the top surface, which is maximally compressed).

For $F <20$ MPa, the distribution of $P_1$ consists of two cup-like counter-polarized closure domains located near the slab ends. The size of the domains increases slightly with an increase in $y'$ from $-B/2$ to $+B/2$. The size of the domain near the clamped end increases significantly with an increase in $F$. For $F <20$ MPa the distribution of $P_3$ consists of two counter-polarized domains, which occupy the central part of the slab. The length of the domains decreases slightly with an increase in $y'$ from $-B/2$ to $+B/2$. The size of the closure domain near the clamped end increases significantly with an increase in $F$.

For $F \geq 20$ MPa, the distribution of $P_1$ splits into domain stripes within the compressed part of the slab ($y' > 0$). These stripes are absent in the stretched part ($y' < 0$) of the slab, regardless of the magnitude of $F$. As $F$ increases above 20 MPa, the number of striped $P_1$-domains and their contrast increase; eventually filling the slab for $F \geq 60$ MPa and $y' > B/4$. At the same time, $P_3$-domains disappear for $F \geq 60$ MPa and $y' > B/4$. The domain structure of $P_3$ is almost independent of the magnitude of $F$ in the stretched part of the slab ($y' < 0$).

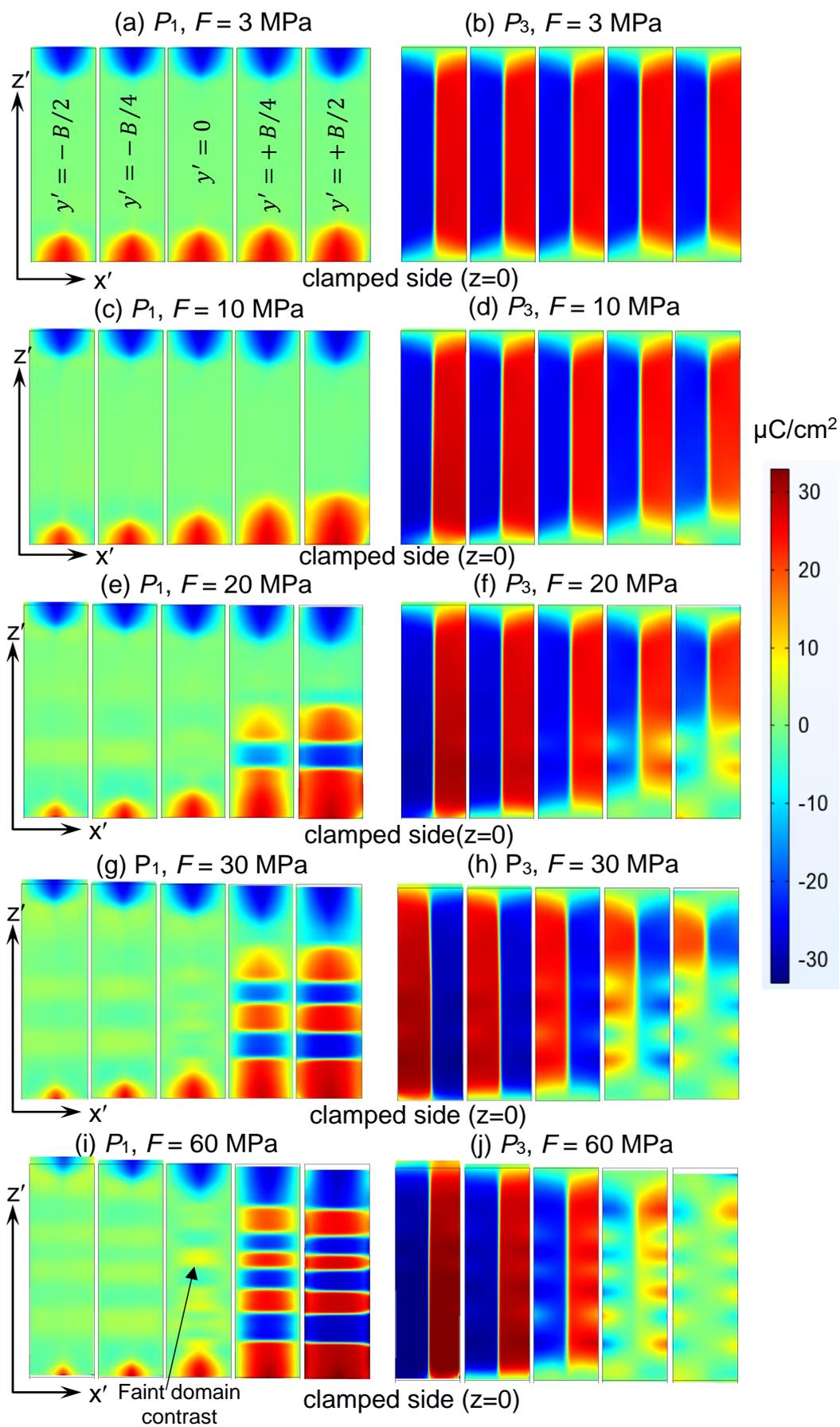

**FIGURE 11**. The distribution of the polarization components $P_1$ (**a, c, e, g, i**) and $P_3$ (**b, d, f, h, j**), calculated

for different cross-sections of a curved BaTiO$_3$ nano-slab with dimensions $4 \times 12 \times 40$ nm$^3$ at room temperature and for the screening length $\lambda =1$ nm. Each group of images consists of five cross-sections perpendicular to the smallest size of the slab, highlighting the bending deformation, which increases with increasing shear force. The images correspond to different depths $y'$, namely, $y' = -B/2, -B/4, 0, B/4$, and $B/2$ (from the left side to the right side of the figure). The slab is clamped at the side $z = 0$ and a shear force with the density $F$ is applied to the opposite side $z' = L$. The force density $F$ is 3 MPa **(a, b)**, 10 MPa **(c, d)**, 20 MPa **(e, f)**, 30 MPa **(g, h)**, and 60 MPa **(i, j)**.

The results presented in **Fig. 11** highlight that it is possible to control the appearance and number of domain stripes in BaTiO$_3$ nano-slabs by varying their curvature with the aid of external shear force applied to the free side of the slab. A highly oriented array of such slabs with one side clamped to the substrate and another side subjected to a shear force can be used for race-track memory elements and/or piezo-nano-generators.

## IV. CONCLUSIONS

It was confirmed that lattice strain depends directly on Ba supersaturation, with higher supersaturation resulting in increased strain of BaTiO$_3$ nanorods synthesized under varying conditions. However, the strain does not directly influence crystal growth or lattice distortion. Instead, OH$^-$ ions incorporation can play a significant role in these processes. The slower nucleation rate, achieved by using the less reactive TiO$_2$ precursor compared to TiOCl$_2$, along with control of Ba supersaturation, enables more effective regulation of OH$^-$ incorporation and crystal growth. This in turn affects both particle size and lattice distortion, leading to a $c/a$ ratio of 1.013 – 1.014. OH$^-$ ions induce lattice elongation along the c-axis, contributing to anisotropic growth, increasing of the rod diameter and their growth-induced bending.

To explore the possibility of curvature-induced changes in BaTiO$_3$ nanorod domain morphology, LGD-based analytical calculations and FEM were employed. These methods revealed the appearance of domain stripes in nanorods curved beyond a critical angle. This curvature-controlled domain stripe formation may find application in flexible race-track memory elements for flexo-tronics and domain-wall electronics. This work enhances the understanding of how the shape anisotropy, lattice strains, and strain gradients influence the domain morphology of ferroelectric nanorods, offering a pathway for tuning properties of the nanorods for advanced applications in nanoelectronics.

**Authors contribution.** O.A.K. synthesized the nanoparticle, performed data analysis (XRD, SEM, strain calculations), and wrote the draft of the corresponding manuscript sections. S.D.S.

performed the XRD and SEM measurements. M.M.K. assisted with the synthesis process by consulting and providing materials and conditions. Z.K. assisted with organizational and collaborative matters. S.D.S., M.M.K., and Z.K. edited the draft of the corresponding manuscript sections. E.A.E. and A.N.M. performed theoretical calculations: E.A.E. wrote the codes and prepared figures, A.N.M. formulated the problem, performed analytical calculations, and wrote the draft of the theory section in the manuscript. Y.O.Z. performed NMR studies and wrote the draft of the corresponding section in the manuscript. L.D. performed HR TEM of the samples. D.R.E. worked on conceptualization, results explanation, comparison of theory and experiment, discussion, conclusions, and manuscript text improvements.

**Acknowledgments.** Nanoparticles preparation and basic (SEM, XRD) characterization of the samples were carried out with the financial support of the European Union's Horizon H2020-MSCA-RISE research and innovation actions 778072–ENGIMA–H2020–MSCA–RISE–2017. Nanoparticles preparation was carried out by O.A.K. with a consultation of M.M.K. XRD and SEM analyses were carried out by S.D.S. and analyzed by O.A.K. Calculation of the sample strain based on XRD data was carried out by O.A.K. and sponsored by the NATO Science for Peace and Security Programme under grant SPS G5980 "FRAPCOM". The work of A.N.M. is funded by the National Research Foundation of Ukraine (project "Manyfold-degenerated metastable states of spontaneous polarization in nanoferroics: theory, experiment, and perspectives for digital nanoelectronics," grant application 2023.03/0132), and by the EOARD project 9IOE063 (related STCU partner project is P751b). The work of E.A.E. and Y.O.Z. are funded by the National Research Foundation of Ukraine (project "Silicon-compatible ferroelectric nanocomposites for electronics and sensors," grant application 2023.03/0127).

## Supplementary Materials
### APPENDIX A. The LGD Free Energy Functional

The LGD free energy functional $G$ of the core polarization $\boldsymbol{P}$ additively includes a Landau expansion on the 2-nd, 4-th, 6-th, and 8-th powers of the polarization, $G_{Landau}$; a polarization gradient energy contribution, $G_{grad}$; an electrostatic contribution, $G_{el}$; the elastic, linear, and nonlinear electrostriction couplings and flexoelectric contributions, $G_{es+flexo}$; and a surface energy, $G_S$. The functional $G$ has the form [37, 38, 39]:

$$G = G_{Landau} + G_{grad} + G_{el} + G_{es+flexo} + G_{VS} + G_S, \quad (A.2a)$$

$$G_{Landau} = \int_{0<r<R_c} d^3r \left[ a_i P_i^2 + a_{ij} P_i^2 P_j^2 + a_{ijk} P_i^2 P_j^2 P_k^2 + a_{ijkl} P_i^2 P_j^2 P_k^2 P_l^2 \right], \quad (A.2b)$$

$$G_{grad} = \int_{0<r<R} d^3r \frac{g_{ijkl}}{2} \frac{\partial P_i}{\partial x_j} \frac{\partial P_k}{\partial x_l}, \tag{A.2c}$$

$$G_{el} = -\int_{0<r<R_c} d^3r \left(P_i E_i + \frac{\varepsilon_0 \varepsilon_b}{2} E_i E_i\right) - \frac{\varepsilon_0}{2} \int_{R_c<r<R_s} \varepsilon_{ij}^S E_i E_j d^3r - \frac{\varepsilon_0}{2} \int_{r>R+\Delta R} \varepsilon_{ij}^e E_i E_j d^3r, \tag{A.2d}$$

$$G_{es+flexo} = -\int_{0<r<R_c} d^3r \left(\frac{s_{ijkl}}{2} \sigma_{ij}\sigma_{kl} + Q_{ijkl}\sigma_{ij}P_k P_l + Z_{ijklmn}\sigma_{ij}P_k P_l P_m P_n + \right.$$
$$\left. \frac{1}{2} W_{ijklmn}\sigma_{ij}\sigma_{kl}P_m P_n + F_{ijkl}\sigma_{ij}\frac{\partial P_l}{\partial x_k}\right), \tag{A.2e}$$

$$G_S = \frac{1}{2}\int_{r=R_c} d^2r\, a_{ij}^{(S)} P_i P_j. \tag{A.2f}$$

The coefficient $a_i$ linearly depends on temperature $T$:

$$a_i(T) = \alpha_T[T - T_C(R_c)], \tag{A.3a}$$

where $\alpha_T$ is the inverse Curie-Weiss constant, and $T_C(R_c)$ is the ferroelectric Curie temperature renormalized by electrostriction and surface tension as [37, 38]:

$$T_C(R_c) = T_C\left(1 - \frac{Q_c}{\alpha_T T_C}\frac{2\mu}{R_c}\right), \tag{A.3b}$$

where $T_C$ is a Curie temperature of a bulk ferroelectric. $Q_c$ is the sum of the electrostriction tensor diagonal components, which is positive for most ferroelectric perovskites with cubic m3m symmetry in the paraelectric phase, namely $0.005 < Q_c < 0.05$ (in m$^4$/C$^2$). $\mu$ is the surface tension coefficient.

**Table AI.** LGD coefficients and other material parameters of a BaTiO$_3$ core in Voigt notations. Adapted from Ref.[20].

| Parameter, its description, and dimension (in the brackets) | The numerical value or variation range of the LGD parameters |
|---|---|
| Expansion coefficients $a_i$ in the term $a_i P_i^2$ in Eq.(A.2b) (C$^{-2}\cdot$mJ) | $a_1 = 3.33(T-383)\times 10^5$ |
| Expansion coefficients $a_{ij}$ in the term $a_{ij}P_i^2 P_j^2$ in Eq.(A.2b) (C$^{-4}\cdot$m$^5$J) | $a_{11} = 3.6\,(T-448)\times 10^6$, $a_{12} = 4.9\times 10^8$ |
| Expansion coefficients $a_{ijk}$ in the term $a_{ijk}P_i^2 P_j^2 P_k^2$ in Eq.(A.2b) (C$^{-6}\cdot$m$^9$J) | $a_{111} = 6.6\times 10^9$, $a_{112} = 2.9\times 10^9$, $a_{123} = 3.64\times 10^{10}+7.6(T-293)\times 10^{10}$. |
| Expansion coefficients $a_{ijkl}$ in the term $a_{ijkl}P_i^2 P_j^2 P_k^2 P_l^2$ in Eq.(A.2b) (C$^{-8}\cdot$m$^{13}$J) | $a_{1111} = 4.84\times 10^7$, $a_{1112} = 2.53\times 10^7$, $a_{1122} = 2.80\times 10^7$, $a_{123} = 9.35\times 10^7$. |

| Linear electrostriction tensor $Q_{ijkl}$ in the term $Q_{ijkl}\sigma_{ij}P_kP_l$ in Eq.(A.2e) (C$^{-2}$·m$^4$) | In Voigt notations $Q_{ijkl} \to Q_{ij}$, which are equal to $Q_{11}$=0.11, $Q_{12}$= –0.045, $Q_{44}$=0.059 |
|---|---|
| Nonlinear electrostriction tensor $Z_{ijklmn}$ in the term $Z_{ijklmn}\sigma_{ij}P_kP_lP_mP_n$ in Eq.(A.2e) (C$^{-4}$·m$^8$) | In Voigt notations $Z_{ijklmn} \to Z_{ijk}$. $Z_{ijk}$ varies in the range $-1 \leq Z_c \leq 1$ [40] |
| Nonlinear electrostriction tensor $W_{ijklmn}$ in the term $W_{ijklmn}\,\sigma_{ij}\sigma_{kl}P_mP_n$ in Eq.(A.2e) (C$^{-2}$·m$^4$ Pa$^{-1}$) | In Voigt notations $W_{ijklmn} \to W_{ijk}$. $W_{ij3}$ varies in the range of $0 \leq W_c \leq 10^{-12}$ as a very small free parameter, and we can neglect it, $W_{ij3}=0$ |
| Elastic compliances tensor, $s_{ijkl}$, in Eq.(A.2e) (Pa$^{-1}$) | In Voigt notations $s_{ijkl} \to s_{ij}$, which are equal to $s_{11}$=8.3×10$^{-12}$, $s_{12}$= –2.7×10$^{-12}$, $s_{44}$=9.24×10$^{-12}$. |
| Polarization gradient coefficients $g_{ijkl}$ in Eq.(A.2c) (C$^{-2}$m$^3$J) | In Voigt notations $g_{ijkl} \to g_{ij}$, which are equal to: $g_{11}$=1.0×10$^{-10}$, $g_{12}$= 0.3×10$^{-10}$, $g_{44}$= 0.2×10$^{-10}$. |
| Surface energy coefficients $a_{ij}^{(S)}$ in Eq.(A.2f) | 0 (that corresponds to the natural boundary conditions) |
| Core radius $R_c$ (nm) | Variable: 5 – 50 |
| Background permittivity $\varepsilon_b$ in Eq.(A.2d) (unity) | 7 |

\* $\alpha = 2a_1$, $\beta = 4a_{11}$, $\gamma = 6a_{111}$, and $\delta = 8a_{1111}$

### APPENDIX B. Fitting Parameters for the BaTiO$_3$ Nanorod Samples

Using the experimentally measured values of the tetragonality $\frac{c}{a}$ and the parameter $\Delta c/c$, the average polarization $\sqrt{\langle P_s^2 \rangle}$ and shell strain $\langle \delta u \rangle$ were calculated for different values of the shell relative volume $\delta V$ based on Eqs.(6) (**Fig. B1**). As seen in **Fig. B1**, the separation between the curves for samples S1, S3, S4, and S5 (represented by black, red, green, and blue lines, respectively) changes only slightly with $\delta V$. The values $\sqrt{\langle P_s^2 \rangle}$ and $\langle \delta u \rangle$ on each of these curves increase monotonically with an increase in $\delta V$ from 0 to 0.5. This indicates the reliability of the fitting procedure. However, the result of the fitting is such that the average polarization is about 46 – 63 µC/cm$^2$, which is almost twice as large as the value of bulk BaTiO$_3$ at room temperature ($P_s$ =24 – 27 µC/cm$^2$), but significantly smaller in comparison with the spontaneous polarization experimentally observed in BaTiO$_3$ nanocubes [18]. The shell strains exceed 10 – 20 %, which are unrealistically large values.

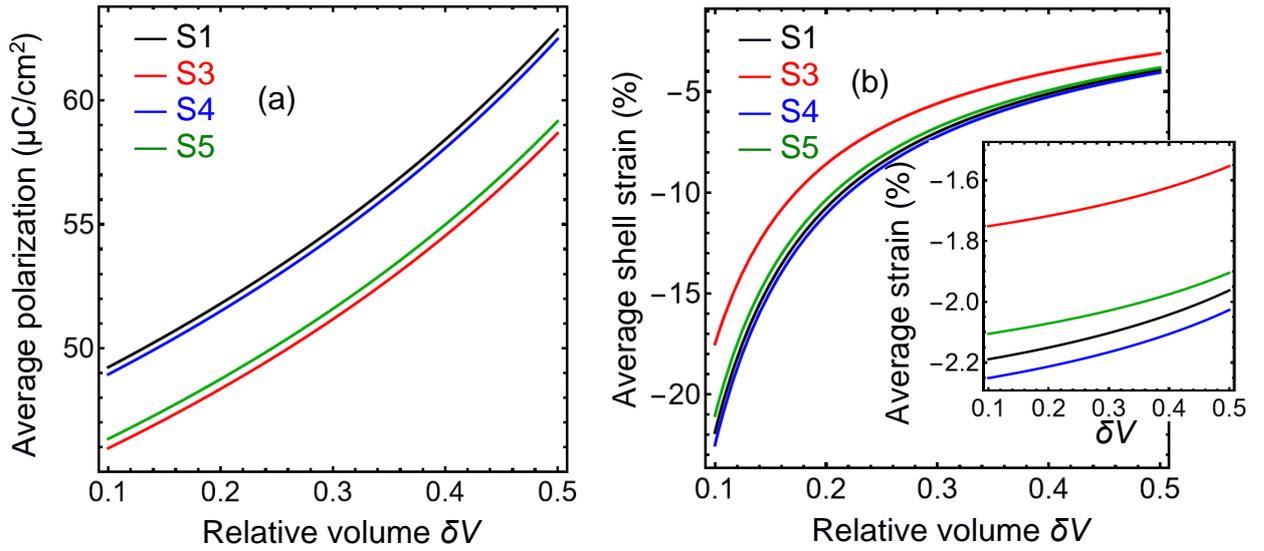

**FIGURE B1.** The average polarization $\sqrt{\langle P_s^2 \rangle}$ (a) and the shell strain $\langle \delta u_s \rangle$ (b) as a function of relative shell volume $\delta V$ obtained from a fit to the experimental results for BaTiO$_3$ nanorod samples S1, S3, S4, and S5 (see legend and **Table I**) at room temperature $T = 298$ K and for $\langle u_c \rangle = 0$. The inset shows the average strain of the nanorod, $\langle \delta u_s \rangle \delta V + \langle u_c \rangle (1 - \delta V)$, where we set $\langle u_c \rangle = 0$.

## APPENDIX C. High Resolution Transmission Electron Microscopy (HR TEM) of the BaTiO$_3$ Nanorods

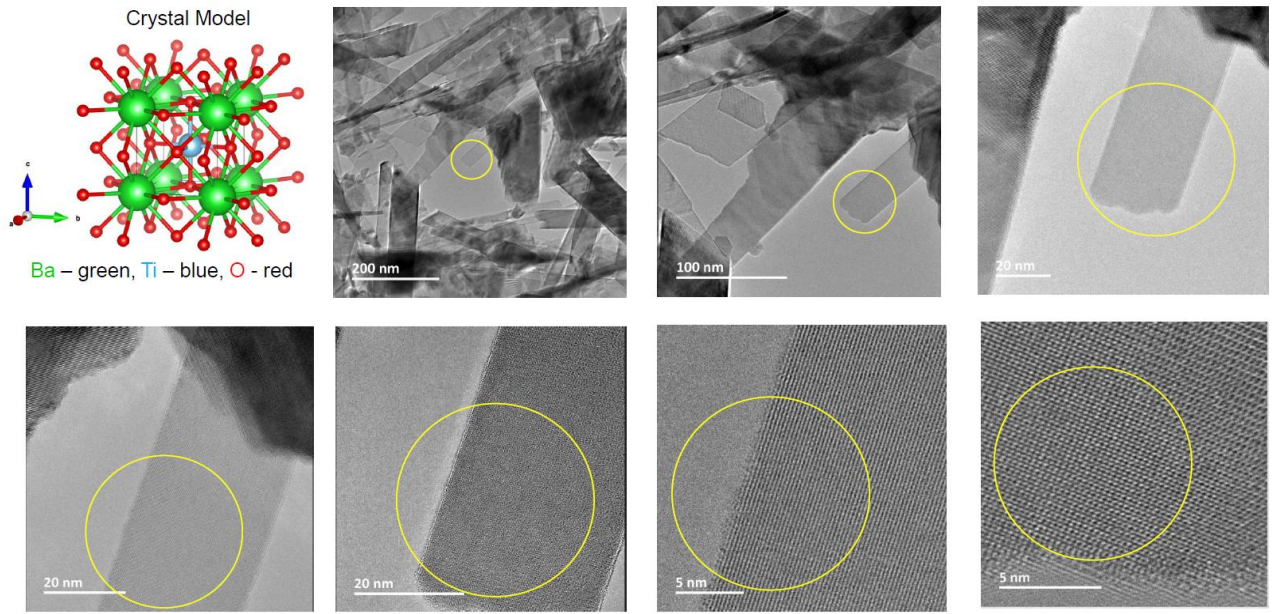

**FIGURE C1.** HR TEM images of the BaTiO$_3$ nanorods (sample S4)

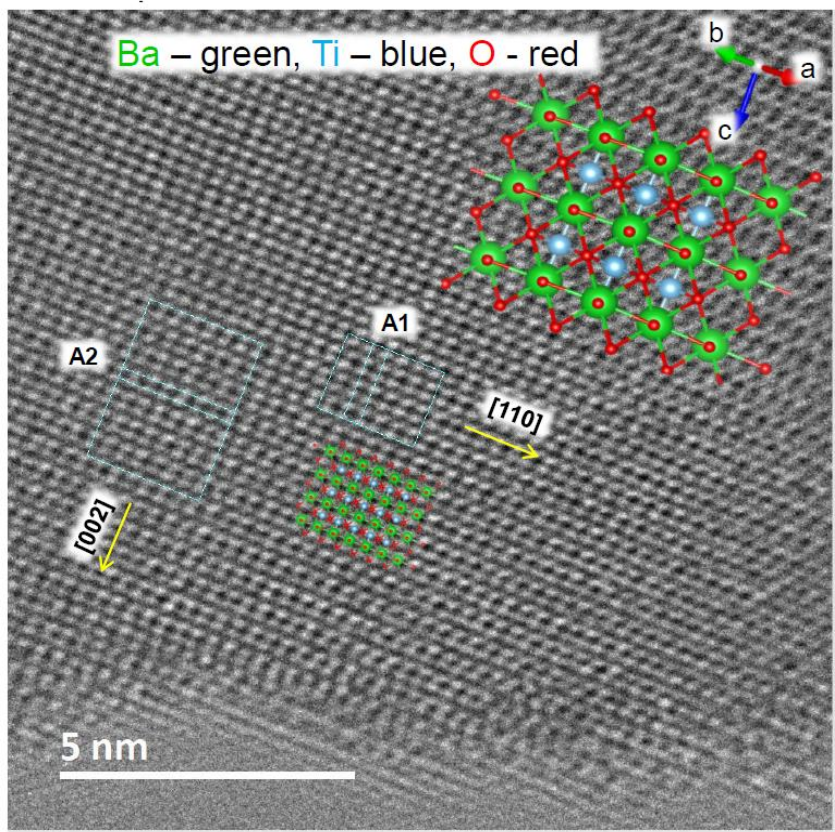

**FIGURE C2.** HR TEM images of the BaTiO$_3$ nanorods (sample S4). The d-spacing in the region A1 (along the long side of the nanorod) is 2.87 Å corresponding to $d_{110}$. The d-spacing in the region A2 (along the short side of the nanoparticle) is 2.05 Å corresponding to $d_{002}$.

# References

[1]   J. Hwang, Z. Feng, N. Charles, X.R. Wang, D. Lee, K.A. Stoerzinger, S. Muy, R.R. Rao, D. Lee, R. Jacobs, D. Morgan, and Y. Shao-Horn. Tuning perovskite oxides by strain: Electronic structure, properties, and functions in (electro) catalysis and ferroelectricity. Materials Today, **31**, 100 (2019); https://doi.org/10.1016/j.mattod.2019.03.014

[2]   H.S. Kim, and N.G. Park. Importance of tailoring lattice strain in halide perovskite crystals. NPG Asia Mater **12**, 78 (2020); https://doi.org/10.1038/s41427-020-00265-w

[3]   S. Dahbi, N. Tahiri, O. El Bounagui, and H. Ez-Zahraouy. Electronic, optical, and thermoelectric properties of perovskite BaTiO$_3$ compound under the effect of compressive strain. Chemical Physics, **544**, 111105 (2021); https://doi.org/10.1016/j.chemphys.2021.111105

[4]   J.D. Ai, C.C. Jin, D.M. Liu, J.T. Zhang, and L.X. Zhang, Strain engineering to boost piezocatalytic activity of BaTiO$_3$. ChemCatChem, **15**, e202201316 (2023); https://doi.org/10.1002/cctc.202201316

[5]   M.J. Choi, J.W. Lee, and H.W. Jang. Strain engineering in perovskites: Mutual insight on oxides and halides. Advanced Materials, **36**, 2308827 (2024); https://doi.org/10.1002/adma.202308827

[6]   U. Acevedo-Salas, B. Croes, Y. Zhang, O. Cregut, K. D. Dorkenoo, B. Kirbus, E. Singh, H. Beccard, M. L. M. Eng, R. Hertel, E. A. Eliseev, A. N. Morozovska, S. Cherifi-Hertel. Impact of 3D curvature on the

polarization orientation in non-Ising domain walls. Nano Letters **23**, 795 (2023); https://doi.org/10.1021/acs.nanolett.2c03579

[7] Y. Liu, A. N. Morozovska, A. Ghosh, K. P. Kelley, E. A. Eliseev, J. Yao, Y. Liu, and S. V. Kalinin. Disentangling stress and curvature effects in layered 2D ferroelectric $CuInP_2S_6$. ASC Nano, **17**, 22004 (2023); https://doi.org/10.1021/acsnano.3c08603

[8] A. N. Morozovska, E. A. Eliseev, Y. Liu, K. P. Kelley, A. Ghosh, Y. Liu, J. Yao, N. V. Morozovsky, A. L Kholkin, Y. M. Vysochanskii, and S. V. Kalinin. Bending-induced isostructural transitions in ultrathin layers of van der Waals ferrielectrics. Acta Materialia, **263**, 119519 (2024); https://doi.org/10.1016/j.actamat.2023.119519

[9] D. Pesquera, K. Cordero-Edwards, M. Checa, I. Ivanov, B. Casals, M. Rosado, J. M. Caicedo, L. Casado-Zueras, J. Pablo-Navarro, C. Magén, J. Santiso, N. Domingo, G. Catalan, F. Sandiumenge, Hierarchical domain structures in buckled ferroelectric free sheets, arXiv preprint arXiv: 2411.19599 (2024); https://doi.org/10.48550/arXiv.2411.19599

[10] G. Segantini, L. Tovaglieri, C. J. Roh, C.-Y. Hsu, S. Cho, R. Bulanadi, P. Ondrejkovic, P. Marton, J. Hlinka, S. Gariglio, D. T.L. Alexander, P. Paruch, J.-M. Triscone, C. Lichtensteiger, A. D. Caviglia. Curvature-Controlled Polarization in Adaptive Ferroelectric Membranes. arXiv preprint arXiv:2503.05452 (2025), https://doi.org/10.48550/arXiv.2503.05452

[11] S. Cherifi-Hertel, C. Voulot, U. Acevedo-Salas, Y. Zhang, O. Crégut, K. D. Dorkenoo, R. Hertel. Polarization-induced topological phase transition in zigzag chains composed of metal nanoparticles J. Appl. Phys., **129**, 243103 (2021); https://doi.org/10.1063/5.0054141

[12] K. J. Choi, M. Biegalski, Y. L. Li, A. Sharan, J. Schubert, R. Uecker, P. Reiche, Y. B. Chen, X. Q. Pan, V. Gopalan, L.-Q. Chen, D. G. Schlom, and C. B. Eom. Enhancement of Ferroelectricity in Strained $BaTiO_3$ Thin Films. Science, **306**, 1005 (2004); https://doi.org/10.1126/science.1103218

[13] C. Ederer and N. A. Spaldin, Effect of epitaxial strain on the spontaneous polarization of thin film ferroelectrics. Phys. Rev. Lett. **95**, 257601, (2005); https://doi.org/10.1103/PhysRevLett.95.257601

[14] M.D. Glinchuk, A.N. Morozovska, E.A. Eliseev. Ferroelectric thin films phase diagrams with self-polarized phase and electret state. J. Appl. Phys. **99**, 114102 (2006); https://doi.org/10.1063/1.2198940

[15] K. P. Kelley, A. N. Morozovska, E. A. Eliseev, V. Sharma, D. E. Yilmaz, A. C. T. van Duin, P. Ganesh, A. Borisevich, S. Jesse, P. Maksymovych, N. Balke, S. V. Kalinin, R. K. Vasudevan. Oxygen vacancy injection as a pathway to enhancing electromechanical responses in ferroelectrics. Adv. Mater. **34**, 2106426 (2021); https://doi.org/10.1002/adma.202106426

[16] S. A. Basun, G. Cook, V. Y. Reshetnyak, A. V. Glushchenko, and D. R. Evans, Dipole moment and spontaneous polarization of ferroelectric nanoparticles in a nonpolar fluid suspension. Phys. Rev. B **84**, 024105 (2011); https://doi.org/10.1103/PhysRevB.84.024105 (Editor's Selection)

[17] D. R. Evans, S. A. Basun, G. Cook, I. P. Pinkevych, and V. Yu. Reshetnyak. Electric field interactions and aggregation dynamics of ferroelectric nanoparticles in isotropic fluid suspensions. Phys. Rev. B, **84,** 174111 (2011); https://doi.org/10.1103/PhysRevB.84.174111

[18]     Yu. A. Barnakov, I. U. Idehenre, S. A. Basun, T. A. Tyson, and D. R. Evans. Uncovering the Mystery of Ferroelectricity in Zero Dimensional Nanoparticles. Royal Society of Chemistry, Nanoscale Adv. **1,** 664 (*2019),* https://doi.org/10.1039/C8NA00131F

[19]     H. Zhang, S. Liu, S. Ghose, B. Ravel, I. U. Idehenre, Y. A. Barnakov, S. A. Basun, D. R. Evans, and T. A. Tyson. Structural Origin of Recovered Ferroelectricity in $BaTiO_3$ Nanoparticles. Phys. Rev. B **108**, 064106 (2023); https://doi.org/10.1103/PhysRevB.108.064106

[20]     E. A. Eliseev, A. N. Morozovska, S. V. Kalinin, and D. R. Evans. Strain-Induced Polarization Enhancement in $BaTiO_3$ Core-Shell Nanoparticles. Phys. Rev. B. **109**, 014104 (2024); https://doi.org/10.1103/PhysRevB.109.014104

[21]     M. Inada, N. Enomoto, K. Hayashi, J. Hojo, S. Komarneni, Facile synthesis of nanorods of tetragonal barium titanate using ethylene glycol, Ceram. Int. 41 (2015) 5581–5587. https://doi.org/10.1016/j.ceramint.2014.12.137 .

[22]     O. Kovalenko, S.D. Škapin, M.M. Kržmanc, D. Vengust, M. Spreitzer, Z. Kutnjak, and A. Ragulya. Formation of single-crystalline $BaTiO_3$ nanorods from glycolate by tuning the supersaturation conditions. Ceramics International, **48**, 11988 (2022); https://doi.org/10.1016/j.ceramint.2022.01.048

[23]     K. Hongo, S. Kurata, A. Jomphoak, M. Inada, K. Hayashi, R. Maezono. Stabilization Mechanism of the Tetragonal Structure in a Hydrothermally Synthesized $BaTiO_3$ Nanocrystal, Inorg. Chem. **57**, 5413 (2018); https://doi.org/10.1021/acs.inorgchem.8b00381

[24]     A. Abragam, *Principles of Nuclear Magnetism* (Oxford University Press, New York, 1961); ISBN 019852014X, 9780198520146

[25]     J. F. Bangher, P. C. Taylor, T. Oja, and P. J. Bray, Nuclear Magnetic Resonance Powder Patterns in the Presence of Completely Asymmetric Quadrupole and Chemical Shift Effects: Application to Matavanadates, J. Chem. Phys. **50**, 4914 (1969); https://doi.org/10.1063/1.1670988

[26]     M. Acosta, N. Novak, V. Rojas, S. Patel, R. Vaish, J. Koruza, G. A. Rossetti, Jr. J. Rödel. $BaTiO_3$-based piezoelectrics: Fundamentals, current status, and perspectives*, Applied Physics Reviews*, **4**, 041305 (2017). https://doi.org/10.1063/1.4990046

[27]     C. Gervais, D. Veautier, M.E. Smith, F. Babonneau, P. Belleville, and C. Sanchez. Solid state $^{47, 49}$Ti, $^{87}$Sr and $^{137}$Ba NMR characterisation of mixed barium/strontium titanate perovskites. Solid State Nuclear Magnetic Resonance, **26** (3-4), 147 (2004); https://doi.org/10.1016/j.ssnmr.2004.03.003

[28]     G. Czjzek, J. Fink, F. Götz, H. Schmidt, J. M. D. Coey, J.-P. Rebouillat, and A. Liénard. Atomic coordination and the distribution of electric field gradients in amorphous solids. Phys. Rev. B **23**, 2513 (1981); https://doi.org/10.1103/PhysRevB.23.2513

[29]     R.H. Meinhold, R.C.T. Slade and R.H. Newman. High Field MAS NMR, with Simulations of the Effects of Disorder on Lineshape. Applied to Thermal Transformations of Alumina Hydrates. Appl. Magn. Reson. **4**, 121 (1993); https://doi.org/10.1007/BF03162559

[30]     J. D. Coster, A. L. Blumenfeld, J. J. Fripiat. Lewis Acid Sites and Surface Aluminum in Aluminas and Zeolites: A High-Resolution NMR Study. Phys. Chem. **98**, 6201, (1994);

https://doi.org/10.1021/j100075a024

[31] T.J. Bastow, H.J. Whitfield. $^{137}$Ba and $^{47,49}$Ti NMR: electric field gradients in the non-cubic phases of BaTiO$_3$. Solid State Communications **117**, 483 (2001); https://doi.org/10.1016/S0038-1098(00)00491-9

[32] G. H. Kwei, A. C. Lawson, S. J. L. Billinge, and S. W. Cheong. Structures of the ferroelectric phases of barium titanate. The Journal of Physical Chemistry. 97 (10), 2368 (1993); https://doi.org/10.1021/j100112a043

[33] O. A. Kovalenko, PhD. Thesis, (in Ukrainian) (2024), https://scholar.google.com/citations?view_op=view_citation&hl=en&user=ogo__V0AAAAJ&sortby=pubdate&citation_for_view=ogo__V0AAAAJ:9ZlFYXVOiuMC

[34] S. V. Kalinin, Y. Kim, D. Fong, and A. Morozovska. Surface Screening Mechanisms in Ferroelectric Thin Films and its Effect on Polarization Dynamics and Domain Structures. Rep. Prog. Phys. **81**, 036502 (2018); https://doi.org/10.1088/1361-6633/aa915a

[35] A. N. Morozovska, E. A. Eliseev, O. A. Kovalenko, and Dean R. Evans. The Influence of Chemical Strains on the Electrocaloric Response, Polarization Morphology, Tetragonality and Negative Capacitance Effect of Ferroelectric Core-Shell Nanorods and Nanowires. Phys. Rev. Applied **21**, 054035 (2024); https://doi.org/10.1103/PhysRevApplied.21.054035

[36] L. D. Landau, and I. M. Khalatnikov. On the anomalous absorption of sound near a second order phase transition point. In Dokl. Akad. Nauk SSSR, **96**, 469 (1954).

[37] E. A. Eliseev, Y. M. Fomichov, S. V. Kalinin, Y. M. Vysochanskii, P. Maksymovich and A. N. Morozovska. Labyrinthine domains in ferroelectric nanoparticles: Manifestation of a gradient-induced morphological phase transition. Phys. Rev. B **98**, 054101 (2018); https://doi.org/10.1103/PhysRevB.98.054101

[38] J. J. Wang, E. A. Eliseev, X. Q. Ma, P. P. Wu, A. N. Morozovska, and Long-Qing Chen. Strain effect on phase transitions of BaTiO$_3$ nanowires. Acta Materialia **59**, 7189 (2011); https://doi.org/10.1016/j.actamat.2011.08.015

[39] A. N. Morozovska, E. A. Eliseev, Y. A. Genenko, I. S. Vorotiahin, M. V. Silibin, Ye Cao, Y. Kim, M. D. Glinchuk, and S. V. Kalinin. Flexocoupling impact on the size effects of piezo- response and conductance in mixed-type ferroelectrics-semiconductors under applied pressure. Phys. Rev. B **94**, 174101 (2016); https://doi.org/10.1103/PhysRevB.94.174101

[40] H. D. Megaw, Temperature changes in the crystal structure of barium titanium oxide. Proceedings of the Royal Society of London. Series A. Mathematical and Physical Sciences **189** (1017), 261 (1947); https://doi.org/10.1098/rspa.1947.0038